\documentclass[11pt, a4paper]{article}

\usepackage[labelfont=bf,textfont=normal]{caption}
 \usepackage[font=small, format=hang, margin={1cm, 1cm}]{caption}
\usepackage[a4paper, left=3cm,right=3cm]{geometry}
\usepackage{amsfonts} 
\usepackage{amsmath} 
\usepackage{mathrsfs}  %
\usepackage{amssymb}
\usepackage{empheq}
\usepackage{mathtools}
\usepackage{slashed}
\usepackage{relsize}
\usepackage{setspace}   
\usepackage{color}  
\usepackage{wasysym}
\usepackage{booktabs}
\usepackage[utf8,applemac]{inputenc} 
\usepackage{tensor}
\usepackage{cite}
\usepackage{tikz}
\usetikzlibrary{calc, arrows.meta}
\usepackage{graphicx}
\usepackage{subfig}
\usepackage{caption}
\usepackage[normalem]{ulem}
\usepackage{hyperref}
\usepackage{epstopdf}
\usepackage{comment}
\usepackage{mathtools}
\newcommand{\D}{\text{d}}

\numberwithin{equation}{section}
\usepackage{dcolumn}
\usepackage{bm}
\usepackage[toc,page]{appendix}

\def\be#1\ee{\begin{align}#1\end{align}}
\def\bsub#1\esub{\begin{subequations}#1\end{subequations}}

\def\f{\frac}

\usepackage{cancel}

\newcommand{\jf}{\varphi}
\newcommand{\pu}{\partial_u}

\def\A{\mathcal{A}}
\def\B{\mathcal{B}}
\def\C{\mathcal{C}}
\def\D{\mathcal{D}}

\def\H{\mathcal{H}}

\def\K{\mathcal{K}}

\def\M{\mathcal{M}}
\def\N{\mathcal{N}}
\def\O{\mathcal{O}}

\def\R{\mathcal{R}}

\def\U{\mathcal{U}}
\def\V{\mathcal{V}}
\def\W{\mathcal{W}}

\def\Y{\mathcal{Y}}
\def\Z{\mathcal{Z}}

\def\de{\mathrm{d}}

\newcommand{\ms}[1]{
    \text{\tiny$#1$}
    }

\newcommand\nts{\negthickspace}
\newcommand\bns{\nts \nts \nts}

\definecolor{bluegreen}{RGB}{0,102,102}

\usepackage{pict2e}
\makeatletter
\newcommand{\loplus}{\mathbin{\mathpalette\dog@lsemi{+}}}
\newcommand{\dog@lsemi}[2]{\dog@semi{#1}{#2}{270,90}}
\newcommand{\dog@semi}[3]{%
  \begingroup
  \sbox\z@{$\m@th#1#2$}%
  \setlength{\unitlength}{\dimexpr\ht\z@+\dp\z@\relax}%
  \makebox[\wd\z@]{\raisebox{-\dp\z@}{%
    \begin{picture}(1,1)
    \linethickness{\variable@rule{#1}}
    \roundcap
    \put(0.5,0.5){\makebox(0,0){\raisebox{\dp\z@}{$\m@th#1#2$}}}
    \put(0.5,0.5){\arc[#3]{0.5}}
    \end{picture}%
  }}%
  \endgroup
}
\newcommand{\variable@rule}[1]{%
  \fontdimen8  
  \ifx#1\displaystyle\textfont3\else
    \ifx#1\textstyle\textfont3\else
      \ifx#1\scriptstyle\scriptfont3\else
        \scriptscriptfont3\relax
  \fi\fi\fi
}
\makeatother

\hypersetup{
    colorlinks,%
    citecolor=blue,%
    filecolor=blue,%
    linkcolor=blue,%
   urlcolor=blue,
   linktoc=page
}

\begin{document}

\title{\Large{\textbf{\sffamily Finite Charges from the Bulk Action}}}
\author{\sffamily Robert McNees${}^1$ and C\'eline Zwikel${}^2$
\date{\small{\textit{
$^1$Department of Physics, Loyola University Chicago
Chicago, IL, USA}\\ rmcnees@luc.edu \\
\textit{$^2$Perimeter Institute for Theoretical Physics,\\ 31 Caroline Street North, Waterloo, ON N2L 2Y5, Canada}\\ czwikel@perimeterinstitute.ca}}}

\maketitle

\begin{abstract}
Constructing charges in the covariant phase space formalism often leads to formally divergent expressions, even when the fields satisfy physically acceptable fall-off conditions. These expressions can be rendered finite by corner ambiguities in the definition of the presymplectic potential, which in some cases may be motivated by arguments involving boundary Lagrangians. We show that the necessary corner terms are already present in the variation of the bulk action and can be extracted in a straightforward way. Once these corner terms are included in the presymplectic potential, charges derived from an associated codimension-2 form are automatically finite. We illustrate the procedure with examples in two and three dimensions, working in Bondi gauge and obtaining integrable charges. As a by-product, actions are derived for these theories that admit a well-defined variational principle when the fields satisfy boundary conditions on a timelike surface with corners. An interesting feature of our analysis is that the fields are not required to be fully on-shell. 
\end{abstract}

\thispagestyle{empty}
\setcounter{page}{1}

\newpage
\tableofcontents
\bigskip
\hrule

\section{Introduction}
The covariant phase space formalism, based on a Lagrangian description of gauge theories, can be used to compute charges associated with 
gauge symmetries \cite{Crnkovic:1986ex,Wald1990OnIC,Lee:1990nz,Iyer:1994ys,Wald:1993nt,Barnich:2001jy,Barnich:2003xg,Barnich:2007bf}, see also the review \cite{Fiorucci:2021pha}. These charges carry physical information about the system and label the states in the phase space. 

In the present work we are interested in theories where the behavior of the fields is specified around a boundary; for instance the spatial boundary. For an asymptotic symmetry $\xi$ that preserves these fall-offs, an associated charge can be defined as the integral of a form $k_{\xi}$ over a closed, codimension-2 surface $\C$ located on this boundary. The Iyer-Wald construction \cite{Lee:1990nz , Iyer:1994ys , Wald:1993nt} can be used to obtain a suitable form $k_{\xi}$, and the candidate charge is then given by 
\begin{gather}\label{eq:BasicChargeDefintion}
    \slashed{\delta} Q_{\xi} = \lim_{r \to \infty} \int_{\C} k_{\xi} ~.
\end{gather}
The limit in this expression, which refers to a coordinate $r$, is shorthand for a procedure involving integrals over a sequence of closed surfaces $\C_{r}$ that approach $\C$.

An immediate problem is that the limit described above may not exist for a given $k_{\xi}$. One is then faced with the problem of somehow adjusting $k_{\xi}$ so that the  $r \to \infty$ limit of its integral over $\C$ exists. A common approach to this problem is to exploit so-called ``corner ambiguities'' in the Iyer-Wald covariant phase space formalism \cite{Lee:1990nz , Iyer:1994ys , Wald:1993nt ,Wald:1999wa}. There, the codimension-2 form satisfies $\de k_{\xi} = \omega_{\xi}$ for a choice of presymplectic current $\omega$ contracted with the symmetry. The presymplectic current is itself obtained from a potential that is defined only up to the addition of a closed form, so there is a natural ambiguity in the choice of $\omega$ and hence the defining property of $k_{\xi}$. The goal, then, is to find an appropriate shift in the codimension-2 form, consistent with this ambiguity, so that the limit \eqref{eq:BasicChargeDefintion} exists.

Once this hurdle is cleared, there is still the issue of the integrability property of $\slashed{\delta}Q_{\xi}$~\footnote{See \cite{Ciambelli_Leigh_2021, Ciambelli_Leigh_2022,Freidel:2021dxw} for another method to discuss the (non-)integrability of charges.}. For theories without propagating degrees of freedom it is expected that the charge is integrable. 
But this can be tedious to show, especially in the presence of ``leaks'' at the boundary \cite{Adami:2020ugu,Adami:2021nnf, Fiorucci:2021pha}. These leaks might be associated with a flux that could pass across the boundary, or simply the fact that there are fields on the boundary with unspecified dynamics, as would be the case for an open system. When there is flux through the boundary, there is a discussion on how to split the charges into integrable and non-integrable parts \cite{Wald:1999wa,Barnich:2011mi,Compere:2018ylh,Adami:2021sko}.

Identifying a corner ambiguity that renders the charges finite can involve trial-and-error, or \emph{ad hoc} justifications that seem to apply only to specific examples.\,\footnote{However, see \cite{Freidel:2019ohg} for a renormalization procedure for electromagnetism for spacetime dimensions higher or equal to six.} But in some cases there are proposals that unambiguously motivate these terms. For example, in Einstein-Hilbert gravity with asymptotically anti-de Sitter boundary conditions, one can construct the charges working in Fefferman--Graham gauge \cite{fefferman1985conformal}. The corner terms used to render the charges finite for various boundary conditions are related to symplectic potentials associated with boundary Lagrangians \cite{ Compere:2008us ,Detournay:2014fva, Compere:2020lrt , Fiorucci:2020xto} used for holographic renormalization of the action \cite{deHaro:2000xn , Bianchi:2001kw}. 
Another proposal for extracting the necessary corner contributions from a boundary Lagrangian was given in \cite{Freidel:2021cjp,Freidel:2021fxf}, though it may be difficult to determine the precise form of the boundary Lagrangian, or if such a Lagrangian exists. Note also that the prescription advocated in \cite{Margalef-Bentabol:2020teu,G:2021xvv,Margalef-Bentabol:2022zso} is inapplicable in presence of leaks.

However, even in a theory with asymptotically AdS boundary conditions, it may be desirable to work in a gauge other than Fefferman-Graham. For example, one might consider AdS spacetimes in Bondi gauge, since this description admits a well-defined flat limit \cite{Barnich:2012aw,Poole:2018koa,Compere:2019bua,Campoleoni:2022wmf}. But when the boundary metric dynamics is fluctuating, it does not seem that the asymptotically AdS construction referenced above can be carried out in Bondi gauge while simultaneously preserving the flat limit \cite{Ruzziconi:2020wrb,Geiller:2021vpg}. Specifically, it was shown that the renormalization of the charges was not connected to the boundary terms renormalizing the action.

In this paper, we show that there is a straightforward way of ensuring that the right-hand side of \eqref{eq:BasicChargeDefintion} exists, using only the variation of the bulk Lagrangian and the fall-offs of the fields at the boundary to identify the necessary corner terms. 
We apply this prescription to two dimensional dilaton gravity theories and three dimensional Einstein gravity (with or without a cosmological constant) with leaky boundary conditions. These lower dimensional theories are widely used as toy models of quantum gravity, tractable examples of black hole evaporation, and concrete examples of holographic dualities \cite{Brown:1986nw, Witten:1988hc, Strominger:1997eq, Grumiller:2002nm, Maldacena:2016hyu, Kitaev:15ur, Sachdev:1992fk, Grumiller:2021cwg, Penington:2019npb, Almheiri:2019psf, Barnich:2012aw,Ecker:2023sua}. The resulting charges \eqref{eq:BasicChargeDefintion} are not only finite, but also symplectic in the sense that the codimension-2 form is independent of the coordinate $r$ used to define the limiting procedure (consistent with previous results obtained in various gauges \cite{Compere:2015knw,Compere:2014cna,Ruzziconi:2020wrb,Geiller:2021vpg,Compere:2017knf,Adami:2023fbm}). With the correct field-dependence assigned to the symmetry parameters, the charges are integrable and agree with or generalize previous results \cite{Ruzziconi:2020wrb,Geiller:2021vpg}.

Our approach ensures that the limiting procedure used to define the charges exists, but a residual ambiguity remains in the finite (as $r \to \infty$) part of the codimension-2 form. However, at least for the lower-dimensional examples considered here, we argue that one prescription for identifying the finite part seems to be more natural in the sense that it leads to integrable charges. In particular, for three-dimensional Einstein gravity in Bondi-Weyl gauge (which is an extension of Bondi gauge that satisfies a weaker condition on the determinant of the transverse metric), this explains the origin of a finite corner term added by hand in \cite{Geiller:2021vpg} to restore integrability of the charges.

The construction of the charges does not reference boundary terms in the action, and there is no requirement that the first variation of the action vanishes (though we do impose a subset of the bulk equations of motion). However, the analysis of the asymptotic structure of the presymplectic potential naturally identifies boundary terms which must be added to the action to obtain a well-defined variational principle for a particular choice of boundary conditions. 
For the two- and three-dimensional theories we consider, we first obtain the charges, and then use our results for the presymplectic potential to work out actions for fields satisfying Dirichlet conditions at the asymptotic boundary. This generalizes results obtained in \cite{Grumiller:2007ju,Ruzziconi:2020wrb,Geiller:2021vpg}. In particular our construction of the action extends to cases where the boundary includes non-smooth corners.

An additional feature of our analysis is that, unlike the usual covariant phase space approach, the fields are not assumed to be fully on-shell. Instead, we enforce a subset of the equations of motion sufficient to determine the $r$-dependence of the fields.~\footnote{In the low-dimensional gravity theories we consider, the $r$-dependence of the fields and their variations is completely fixed. In theories with local degrees of freedom, it will only be determined to some order in a large-$r$ expansion.} The other equations of motion, which are conjugate to fields fixed by our choice of gauge and would normally be imposed as constraints, are not enforced. Nevertheless, we show there is a (not unique) codimension-2 form $k_{\xi}$ sourced by a combination of the presymplectic current and the weakly-vanishing Noether current associated with diffeomorphism invariance. Although the focus here is lower-dimensional theories, this result is completely general and holds for any diffeomorphism invariant theory in any number of spacetime dimensions. It reduces to the usual definition of the codimension-2 form when the fields are fully on-shell, but supports additional charges for our slightly weaker condition. This is relevant, for example, in Jackiw-Teitelboim dilaton gravity \cite{Jackiw:1984, Teitelboim:1984}, which is dual to a quantum mechanical model that exhibits on-shell breaking of conformal symmetry \cite{Maldacena:2016hyu, Kitaev:15ur, Jensen:2016pah, Maldacena:2016upp,Godet:2020xpk,Grumiller:2015vaa}. There, relaxing one of the equations of motion results in charges that realize a Virasoro algebra, but the conformal symmetry is broken when the fields are fully on-shell.

\subsubsection*{Organization of the paper}

In section \ref{sec1:covphasespace} we state our main result and present the proof.  Sections \ref{sec:2d} and \ref{sec:3d} illustrate our procedure applied to two dimensional dilaton gravity in linear dilaton Bondi gauge, and on three dimensional gravity in Bondi-Weyl gauge, respectively. In both cases, we follow up an analysis the charges with the construction of an action principle using our adapted choice of presymplectic potential.

Some details of the calculations in sections \ref{sec:2d} and \ref{sec:3d} have been place in appendices. A final appendix \ref{app:3dFG} translates some of our results for three dimensional gravity back to Fefferman-Graham gauge. 

\subsubsection*{Notation}
In general we use Greek letters $\mu, \nu, \ldots$ for indices on the full two- or three-dimensional spacetime, and letters $a, b, \ldots$ from the beginning of the Roman alphabet for indices on codimension-1 surfaces at finite $r$ or on the asymptotic boundary. A generic index $i$ is used to label the field content. Other notation is explained as it is introduced.

\section{Covariant phase space and charges}\label{sec1:covphasespace}

In the first subsection we give a rough overview of the main result. This is derived in more detail, with a slightly weaker set of assumptions, in the second subsection. Finally, we comment on the relation between our results and the boundary terms needed for an action with a well-defined variational principle.

\subsection{Sketch of main result}

It is easy to see why the $r \to \infty$ limit in the definition \eqref{eq:BasicChargeDefintion} of the charge $\slashed{\delta}Q_{\xi}$ might not exist. Consider a theory on a spacetime $M$ and some choice of presymplectic current $\omega^{\mu}(\Psi; \delta_1 \Psi, \delta_2 \Psi)$. For now we assume that the fields and field variations satisfy the equations of motion, so $\omega$ is closed. In the covariant phase space formalism, the codimension-2 form for an asymptotic symmetry $\xi$ satisfies 
\begin{gather}
	\partial_{\nu} k_{\xi}^{\mu\nu}   = \omega_{\xi}^{\mu}~ 
\end{gather}
where the equality holds on-shell. The right-hand side of this equation is shorthand for the presymplectic current with one of the field variations given by the action of $\xi$ on the fields. Working in coordinates where the boundary is the $r \to \infty$ limit of a constant $r$ surface $\B$, let $\N$ be a surface that intersects $\B$ at a closed, codimension-2 surface $\C$. On a neighborhood of $\B$ the surface $\N$ is an isosurface of some other coordinate $u$, and the charge associated with $\xi$ is then obtained as
\begin{gather}\label{eq:IyerWaldCharge}
    \slashed{\delta} Q_{\xi} = \lim_{r \to \infty}\int_{\C} k_{\xi}^{ur} ~.
\end{gather}
Denoting coordinates on $\C$ by $x^{A}$, this component of the codimension-2 form satisfies
\begin{gather}\label{eq:FiniteChargeObstruction}
    \partial_{r} k_{\xi}^{ur} + \partial_{A} k_{\xi}^{uA} = \omega_{\xi}^{u} ~.
\end{gather}
If $\omega_{\xi}^{u}$ does not fall off fast enough for large $r$, then the integral over $\C$ of $k_{\xi}^{ur}$ will diverge in the limit $r \to \infty$, and the charge is not defined.\,\footnote{One might argue that this means the boundary conditions must be tightened, so that the fields and their variations in $\omega_{\xi}^{u}$ fall off more rapidly with $r$. Our assumption is that the boundary conditions necessary for the theory to admit a sufficiently wide array of solutions have already been determined, and faster fall off in $r$ would be too restrictive.} Working at large but finite $r$, one typically finds that $k_{\xi}^{ur}$ can be written as a part that diverges, a part that remains finite, and terms that vanish in this limit. Even if one could simply discard the divergent part, the finite piece that remains may not describe the expected charge \cite{Adami:2021sko,Geiller:2021vpg}.

To address these problems we take advantage of a natural ambiguity in the Iyer-Wald formalism to identify a choice of presymplectic current such that \eqref{eq:IyerWaldCharge} is well-defined. This choice is not unique; a residual ambiguity will remain. Once the boundary conditions are set, the behavior of the fields is known to some order in a large-r expansion near $\B$. Then the components of the presymplectic potential along $\B$ can be integrated, allowing us to write 
\begin{equation}\label{eq:DerivativeOfIntegral}
 \Theta^a(\Psi;\delta\Psi)=\partial_r\int dr \, \Theta^a(\Psi;\delta\Psi) ~,
\end{equation} 
with $x^a=(u,x^A)$ coordinates on $\B$. Here, the integral is meant to capture the parts of $\Theta^{a}$ common to all field configurations -- there may also be subleading terms associated with degrees of freedom not fixed by the boundary conditions. Expressing the components $\Theta^{a}$ as total $r$-derivatives, the bulk divergence of the presymplectic potential can now be written
\begin{align}\label{eq:definitionThetaren}
\partial_\mu\Theta^\mu(\Psi;\delta\Psi)=\partial_r\left(\Theta^r(\Psi;\delta\Psi)+ \partial_a \int dr \, \Theta^a(\Psi;\delta\Psi)\right) &=:\partial_r\Theta^r_{\text{ren.}}(\Psi;\delta\Psi) \\
\text{ with } \Theta^a_{\text{ren.}}(\Psi;\delta\Psi)& \simeq 0~
\end{align}
Essentially, we shift components of the presymplectic potential along $\mathcal{B}$ to zero (with $\simeq$ indicating the possibility of subleading terms) while preserving their contributions to $\partial_{\mu}\Theta^{\mu}$ as ``corner terms'' in a new $r$-component. This shift in the potential changes both the presymplectic current and the associated codimension-2 form. For the lower-dimensional examples we consider in this paper, $\Theta^{u}$ is the same for all field configurations and $\Theta_\text{ren}^{u} = 0$ with no subleading terms. Then  $\omega_\text{ren}^{u}$ is exactly zero and the integral of \eqref{eq:FiniteChargeObstruction} over $\mathcal{C}$ gives 
\begin{equation}
\partial_r \int_\C k^{ur}_\xi+ \int_\C \partial_Ak^{uA}_\xi = 0 ~.
\end{equation}
The surface $\C$ is closed, so the second term vanishes, and hence the $r \to \infty$ limit exists in the definition of the charge \eqref{eq:IyerWaldCharge}. 
The codimension-2 form obtained from the new presymplectic current satisfies
\begin{gather}\label{eq:onshellddk}
	\partial_{r} \partial_{a} k_{\xi}^{ra} \simeq 0  ~,
\end{gather}
such that $\partial_{a} k_{\xi}^{ra}$ is independent of $r$. For theories with dynamical degrees of freedom the statements above may apply only up to sub-leading terms in a large-$r$ expansion near $\mathcal{B}$. This is discussed in the conclusion.

This is the main result of this paper: We give a prescription to obtain finite charges based solely on corner terms identified in the bulk presymplectic potential. The result does not rely on a particular choice of gauge, or the presence of a particular set of boundary terms in the action.

Note that the conditions above do not completely fix the ambiguity in the definition of the presymplectic potential in the Iyer-Wald formalism. In passing from \eqref{eq:DerivativeOfIntegral} to \eqref{eq:definitionThetaren}, one could still shift $\Theta^{r}_\text{ren.}$ by an $r$-independent piece. However, for the theories we consider, which do not have local degrees of freedom in the bulk, a suitable choice of gauge naturally gives the relevant components of $\Theta^{a}$ as total $r$-derivatives of the fields and their variations. These expressions give corner terms in $\Theta^{r}_\text{ren.}$ with both $r$-dependent and $r$-independent parts, and result in integrable charges $\delta Q_{\xi}$. Thus, for these theories, corner contributions extracted from the bulk presymplectic potential are all that is needed to ensure that the construction in \eqref{eq:IyerWaldCharge} both exists and gives integrable charges.

\subsection{Proof of finite charges at the boundary}
\label{sec1:covphasespacedetail}

Let us now give a more detailed derivation of the results sketched out above. The setting is a diffeomorphism-invariant theory on an $n+1$-dimensional spacetime $M$, describing a collection of tensor fields $\Psi_{i}$ that includes the metric $g$. The definition of the theory specifies boundary conditions for the fields, expressed as conditions on their behavior with respect to some coordinate $r$ as $r \to \infty$. These conditions may refer to quantities that vanish or diverge in this limit, so we work on a spacetime $M$ with a regulating boundary $\partial M$ that includes a component $\B$ which is an isosurface of $r$. Calculations are carried out at large but finite $r$, with the limit $r \to \infty$ taken at the end. (This final step may be implicit in some our results, and we routinely use $\B$ to refer to both the surface at finite $r$ and the component of the boundary at $r \to \infty$.) The boundary may have other components, as well, which possibly intersect $\B$ at corners that comprise $\partial \B$. The fact that $\B$ may have a boundary $\partial \B$ requires careful attention later on when dealing with total derivatives appearing in boundary integrals.

The theory is described by a variational principle that considers all field configurations satisfying the boundary conditions, on manifolds $M$ with a boundary  that includes $\B$ with corners $\partial\B$. These field configurations are weighted by an action functional of the form 
\begin{gather}\label{eq:GenericActionWithBoundary}
   \Gamma = \int_{M} \!\! \de^{n+1}x \, L_{M} + \int_{\partial M}\!\!\! \de^{n}x \, L_{\partial M} ~.
\end{gather}
The boundary Lagrangian $L_{\partial M}$, which may have support on some components of $\partial M$ and not others, will not be relevant for our results on charges. Everything that is needed comes from the bulk part of the action, which changes under an infinitesimal variation of the fields as
\begin{gather}\label{eq:FirstVariationBulkLagrangian}
    \delta L_{M} = E^{i}\,\delta \Psi_{i} + \partial_{\mu}\Theta^{\mu}(\Psi;\delta \Psi) ~.
\end{gather}
The tensor densities $E^{i}$ are the equations of motion and $\Theta^{\mu}$ is the presymplectic potential. Now consider a second variation of the fields. Without loss of generality these field variations can be taken to be independent, so they commute. It follows that the presymplectic current $\omega^{\mu}$, defined as 
\begin{gather}
    \omega^{\mu}(\Psi;\delta_1 \Psi, \delta_2 \Psi) = \delta_{2}\Theta^{\mu}(\Psi;\delta_1\Psi) - \delta_{1}\Theta^{\mu}(\Psi;\delta_2\Psi) ~,
\end{gather}
satisfies
\begin{gather}\label{eq:OffShellSymplecticCurrent}
    \partial_{\mu}\omega^{\mu} + \delta_{2} E^{i}\,\delta_{1} \Psi_{i} - \delta_{1} E^{i}\,\delta_{2}\Psi_{i} = 0 ~.
\end{gather}
This is just a consequence of $[\delta_2, \delta_1] = 0$ applied to \eqref{eq:FirstVariationBulkLagrangian}; it holds independent of the equations of motion. The presymplectic current is conserved, $\partial_{\mu}\omega^{\mu} = 0$, when we restrict to field variations that satisfy the equation of motion.

Under a diffeomorphism $x^\mu \to x^\mu + \xi^\mu$ the fields $\Psi_{i}$ change by their Lie derivative along $\xi$. The response of the bulk Lagrangian is
\begin{gather}
    \delta_{\xi} L_{M} = E^{i}\,\delta_{\xi} \Psi_{i} + \partial_{\mu}\Theta^{\mu}(\Psi;\delta_{\xi} \Psi) .
\end{gather}
After integration-by-parts this can be written as
\begin{gather} \label{eq:derivation Noether id}
    \delta_{\xi} L_{M} = \xi^{\mu} N_{\mu} + \partial_{\mu}\Big(\Theta^{\mu}(\Psi;\delta_{\xi}\Psi) + S^{\mu}_{\xi}\Big) .
\end{gather}
The quantity $N_{\mu}$ -- a combination of the equations of motion, the fields, and their derivatives -- vanishes as a consequence of diffeomorphism invariance. These Noether identities take the form
\begin{gather}\label{eq:NoetherIdentity}
    N_{\mu} = E^{i} \partial_{\mu}\Psi_{i} - \partial_{\nu} \big(E^i \Psi_i\big)^{\nu}{}_{\mu} = 0 ~,
\end{gather}
where $\big(E^i \Psi_i\big)^{\nu}{}_{\mu}$ denotes the contractions of the tensor densities $E^{i}$ and tensors $\Psi_{i}$ appearing in 
\begin{gather}
    E^{i}\,\delta_{\xi}\Psi_{i} = E^{i}\,\xi^{\mu}\partial_{\mu}\Psi_{i} + \big(E^i \Psi_i\big)^{\mu}{}_{\nu} \,\partial_{\mu}\xi^{\nu} ~.
\end{gather}
The current $S^{\mu}_{\xi}$ appearing as a divergence in \eqref{eq:derivation Noether id} is then
\begin{gather}
    S^{\mu}_{\xi} = \xi^{\nu}\,(E^{i}\,\Psi_{i})^{\mu}{}_{\nu} ~.
\end{gather}
This is the ``weakly vanishing Noether current''. It is zero when the fields are fully on-shell -- when all the equations of motion $E^{i}=0$ are satisfied -- but may have non-zero components if one or more of the equations of motion are not enforced.

Diffeomorphism invariance implies $N_{\mu} = 0$ for all field configurations, independent of whether the equations of motion hold. Thus, given two sets of fields that differ by infinitesimal $\delta\Psi_i$, the difference $\delta N_{\mu} = N_{\mu}(\Psi + \delta \Psi) - N_{\mu}(\Psi)$ must also vanish. Linearizing \eqref{eq:NoetherIdentity}, we have
\begin{gather}
    \delta N_{\mu} = \delta E^{i}\,\partial_{\mu} \Psi_{i} + E^{i}\partial_{\mu}\delta \Psi_{i} - \partial_{\nu} \delta \big(E^i \Psi_i\big)^{\nu}{}_{\mu} = 0 ~.
\end{gather}
Contracting this with $\xi^{\mu}$, and recalling that the $E^{i}$ transform as tensor densities, leads to the important identity
\begin{gather}
    \delta E^{i}\,\delta_{\xi}\Psi_{i} - \delta_{\xi} E^{i}\,\delta \Psi_{i} = \partial_{\mu}\Big( \xi^{\nu}\,\delta \big(E^i \Psi_i\big)^{\mu}{}_{\nu} - \xi^{\mu}\,E^{i}\,\delta \Psi_{i} \Big) ~.
\end{gather}
The left-hand side takes the same form as the terms appearing in \eqref{eq:OffShellSymplecticCurrent}, for a generic field variation $\delta_2 = \delta$ and a field variation $\delta_1 = \delta_\xi$ with the same form as a diffeomorphism. Combining these results gives
\begin{gather}\label{eq:OffShellSymplecticCurrentDiffeo}
    \partial_{\mu}\Big(\omega^{\mu}_{\xi} + \xi^{\nu}\,\delta \big(E^{i}\Psi_{i})^{\mu}{}_{\nu} - \xi^{\mu}\, E^{i}\,\delta\Psi_{i}\Big) = 0 ~,
\end{gather}
where $\omega^{\mu}_{\xi} = \omega^{\mu}(\Psi; \delta_{\xi}\Psi; \delta \Psi)$. This equation holds off-shell, like the other results above, and is straightforward to verify for specific models. Since the divergence \eqref{eq:OffShellSymplecticCurrentDiffeo} vanishes, it is always possible to find a (non-unique) codimension-2 form with components $k^{\mu\nu}_{\xi}$ that satisfy
\begin{gather}\label{eq:FirstCodimension2Definition}
    \partial_{\nu}k^{\mu\nu}_{\xi} = \omega^{\mu}_{\xi} + \xi^{\nu}\,\delta \big(E^{i} \Psi_{i})^{\mu}{}_{\nu} - \xi^{\mu}\, E^{i}\,\delta\Psi_{i} ~.
\end{gather}
This is a consequence of diffeomorphism invariance that holds for any field configuration, not just solutions of the equations of motion.\,\footnote{See the discussion around equation (2.2.67) in \cite{Fiorucci:2021pha} for a similar result.}  When the fields are fully on-shell it reduces to the Iyer-Wald definition.

Covariant phase space methods impose the equations of motion $E^{i} = 0$ for the fields. Here, we take a different approach and enforce the slightly weaker condition $E^{i}\,\delta\Psi_{i} = 0$. This can be accomplished by gauge fixing components of some of the fields, restricting to field variations that preserve this gauge, and enforcing only some of the equations of motion. For example, the Bondi gauge used in the next few sections fixes a particular component $g_{rr}$ of the spacetime metric. Field variations that preserve this gauge satisfy $\delta g_{rr} = 0$. As a result, the term $E^{rr}\,\delta g_{rr}$ appearing in $E^{i}\,\delta \Psi_{i}$ will vanish even if the component of the equation of motion conjugate to $g_{rr}$ (which would normally be enforced as a constraint) does not vanish: $E^{rr} \neq 0$. Field configurations that satisfy 
\begin{gather}\label{eq:PartiallyOnShell}
    E^{i}\,\delta\Psi_{i} = 0 
\end{gather}
in this way, through a combination of equations of motion and gauge conditions, will be referred to as ``partially on-shell.'' For partially on-shell fields, the defining equation \eqref{eq:FirstCodimension2Definition} for the codimension-2 form reduces to
\begin{gather}\label{eq:SecondCodimension2Definition}
    \partial_{\nu}k^{\mu\nu}_{\xi} = \omega^{\mu}_{\xi} + \xi^{\nu}\,\delta \big(E^{i} \Psi_{i})^{\mu}{}_{\nu} ~.
\end{gather}
The motivation for imposing \eqref{eq:PartiallyOnShell} rather than the stronger condition $E^{i} = 0$ comes from the last term in this equation. Now the variation of the weakly vanishing Noether current may have non-zero components that contribute to the codimension-2 form, supporting charges that are not present when $E^{i} = 0$. Such charges appear, for instance, in gravitational theories dual to quantum mechanical models that exhibit a larger symmetry algebra when off-shell, which is then broken to something simpler when the fields are fully on-shell \cite{Maldacena:2016hyu, Maldacena:2016upp, Engelsoy:2016xyb, Grumiller:2017qao}.

There are natural ambiguities in the definitions of both $\Theta^{\mu}$ and $\omega^{\mu}$ which may or may not affect the defining property \eqref{eq:SecondCodimension2Definition} of the codimension-2 form. For example, boundary terms in the action can be thought of as shifting the presymplectic potential by a $\delta$-exact contribution; i.e. $\Theta \to \Theta + \delta L_{\partial M}$. However, such a shift has no effect on $\omega^{\mu}$, and a boundary term in the action does not change the equations of motion, so the result \eqref{eq:FirstCodimension2Definition} is not sensitive to this sort of ambiguity in the definition of $\Theta$.\,\footnote{A shift of this sort \textit{is} relevant for the variational problem, as we will discuss below.} On the other hand, it is clear from \eqref{eq:FirstVariationBulkLagrangian} that the presymplectic potential is only defined up to a shift $\Theta^{\mu} \to \Theta^{\mu} + \partial_{\nu}\Upsilon^{\mu\nu}$ for an anti-symmetric $\Upsilon^{\mu\nu}(\Psi;\delta\Psi)$ that is linear in the field variations. This \emph{does} change $\omega^{\mu}$, by a term
\begin{gather}
    \omega^{\mu}(\Psi;\delta_1\Psi; \delta_2\Psi) \to \omega^{\mu}(\Psi;\delta_1\Psi; \delta_2\Psi) + \partial_{\nu}\Y^{\mu\nu}(\Psi;\delta_1\Psi, \delta_{2}\Psi) ~,
\end{gather}
where
\begin{gather}
    \Y^{\mu\nu}(\Psi;\delta_1\Psi, \delta_{2}\Psi) = \delta_{2}\Upsilon^{\mu\nu}(\Psi;\delta_1\Psi) - \delta_{1}\Upsilon^{\mu\nu}(\Psi;\delta_2\Psi) ~.
\end{gather}
The shift in $\omega^{\mu}$ is consistent with both \eqref{eq:OffShellSymplecticCurrent} and \eqref{eq:OffShellSymplecticCurrentDiffeo}, but it contributes a new term to \eqref{eq:FirstCodimension2Definition} that changes $k^{\mu\nu}_{\xi}$ by
\begin{gather}\label{eq:Codimension2FormShift}
    k^{\mu\nu}_{\xi} \to k^{\mu\nu}_{\xi} + \Y^{\mu\nu}(\Psi;\delta_{\xi}\Psi, \delta \Psi) ~.
\end{gather}
For the models we consider, there is a natural prescription for $\Upsilon^{\mu\nu}$, or equivalently $\Y^{\mu\nu}$, such that the resulting codimension-2 form gives finite charges at the boundary.

Consider now the component $\B$ of $\partial M$, defined as a surface of constant $r$. Coordinates along this surface are denoted by $x^{a}$. Since we know the fall-offs of the fields close to the boundary \footnote{ For instance in Bondi gauge these can be obtained once we partially solve the equations of motion.}, the components of the presymplectic potential along $\B$ can be written as total derivatives with respect to $r$: 
\begin{gather}\label{eq:ThetaTotalDerivative}
    \Theta^{a} \simeq \partial_{r} Y^{ar} ~.
\end{gather}
This is trivially true once the $r$-dependence of the fields is known; simply integrating $\Theta^{a}$ gives $Y^{ar}$ up to an $r$-independent ambiguity. However, for the theories we consider, some of the components can already be put in the form \eqref{eq:ThetaTotalDerivative} simply by evaluating $\Theta^{a}$ in Bondi gauge and applying integration-by-parts. For those components, the resulting $Y^{ar}$ are functions of the spacetime fields and their variations, and give specific $r$-dependent and $r$-independent parts when the fields are partially on-shell. Making a shift $\Theta^{\mu} \to \Theta^{\mu} + \partial_{\nu}\Upsilon^{\mu\nu}$ and choosing $\Upsilon^{\mu\nu}$ to set the components $\Theta^{a}$ to zero, gives $\Upsilon^{ar} = - Y^{ar}$ up to the residual ambiguity mentioned above. As a result, we have
\begin{gather}
    \Theta^{r} \to \Theta^{r} + \partial_{a} Y^{ar} \vphantom{\Big|}\\
    \omega^{a}(\Psi;\delta_1\Psi; \delta_2\Psi) \to 0 \vphantom{\Big|} \\
    \omega^{r}(\Psi;\delta_1\Psi; \delta_2\Psi) \to \omega^{r}(\Psi;\delta_1\Psi; \delta_2\Psi) + \partial_{a}\Big(\delta_{2}Y^{ar}(\Psi;\delta_1\Psi) - \delta_{1}Y^{ar}(\Psi;\delta_2\Psi) \Big) ~,
\end{gather}
along with a corresponding shift \eqref{eq:Codimension2FormShift} to the codimension-2 form. Note that both $\Theta^{r}$ and $\omega^{r}$, which naturally appear in integrals over the constant $r$ surface $\B$, acquire corner terms (total derivatives on $\B$) with support on $\partial \B$.

There is one additional modification to the codimension-2 form coming from the variation of the weakly vanishing Noether current in \eqref{eq:SecondCodimension2Definition}. For the theories we consider, its components along $\B$ for an asymptotic symmetry $\xi$ can also be written as total $r$-derivatives
\begin{gather}
    \xi^{\nu}\,\delta\big(E^{i}\Psi_{i}\big)^{a}{}_{\nu} = \partial_{r}\Z^{ar}_{\xi} ~.
\end{gather}
As a result, \eqref{eq:OffShellSymplecticCurrentDiffeo} can be expressed entirely as a total $r$-derivative  
\begin{gather}\label{eq:IndependentOfr}
    \partial_{r}\Big(\omega^{r}_{\xi} + \partial_{a}\Y^{ar}_{\xi} + \xi^{\nu}\,\delta \big(E^{i}\Psi_{i}\big)^{r}{}_{\nu} + \partial_{a}\Z^{ar}_{\xi}\Big) = 0 ~,
\end{gather}
where $\Y^{ar}_{\xi}$ denotes the shift in \eqref{eq:Codimension2FormShift}. This gives a family of codimension-2 forms $k^{\mu\nu}_{\xi}$ that satisfy
\begin{gather}
	\label{eq:partialkEquation}
    \partial_{a} k^{ra}_{\xi} = \omega^{r}_{\xi} + \partial_{a}\Y^{ar}_{\xi} + \xi^{\nu}\,\delta \big(E^{i}\Psi_{i}\big)^{r}{}_{\nu} + \partial_{a}\Z^{ar}_{\xi} ~.
\end{gather}
The quantity $\partial_{a} k^{ra}_{\xi}$ is independent of $r$, by virtue of \eqref{eq:IndependentOfr}. Compared to the codimension-2 form obtained from \eqref{eq:FirstCodimension2Definition}, it is shifted by
\begin{gather}\label{eq:kShifted}
    k^{ra}_{\xi} \to k^{ra}_{\xi} + \Y^{ar}_{\xi} + \Z^{ar}_{\xi} ~,
\end{gather}
up to a residual ambiguity which is independent of $r$.

In the lower-dimensional theories we investigate, setting $\Theta^{u} \to \Theta_\text{ren}^{u} = 0$ via the Iyer-Wald ambiguity described above implies $\omega^{u} \to \omega_\text{ren}^{u} = 0$. The codimension-2 form then satisfies
\begin{gather}\label{eq:partialkEquation2}
    \partial_{r} k^{ur}_{\xi} + \partial_{A} k^{uA}_{\xi} = 0 ~,
\end{gather}
so that the $r \to \infty$ limit of $k_{\xi}^{ur}$ integrated over $\C$ exists. Alternately, one can begin with the right-hand side of \eqref{eq:partialkEquation}, which is explicitly independent of $r$ by virtue of \eqref{eq:IndependentOfr}, and use integration by parts to obtain $k_{\xi}^{ur}$ up to total derivatives on $\C$. The result is not simply the finite part of the problematic codimension-2 form from our earlier statement of the problem; the new terms in \eqref{eq:kShifted} have also contributed terms independent of $r$. Accounting for these finite shifts, we will find that the charges are now integrable once proper field-dependence is assigned to the generators of the asymptotic symmetries. Thus, we have obtained a codimension-2 form such that the construction \eqref{eq:IyerWaldCharge} of the charge associated with the symmetry $\xi$ exists and is integrable, using only quantities arising in the variation of the bulk Lagrangian.

This result can be generalized to anomalous Lagrangians, and to state dependent asymptotic symmetries following the mathematical framework developed for instance in \cite{Fiorucci:2020xto,Freidel:2021cjp}. {It can also be generalized for fields transforming with higher derivatives of the symmetry parameters, see for instance \cite{Compere:2007az}. }

\subsection{A comment on boundary terms and variational principles}
\label{sec12:VariationalPrinciples}

The construction detailed above makes no reference to specific boundary terms in the action \eqref{eq:GenericActionWithBoundary}. But it is relevant for identifying boundary terms that must be added to the bulk action to obtain a well-defined variational principle for a given set of boundary conditions, including cases where $\B$ is taken to have a boundary $\partial \B$.

When the fields are partially on-shell ($E^{i}\delta\Psi_{i} = 0$) or fully on-shell ($E^{i} = 0$), the divergence of the presymplectic potential is the bulk contribution to the variation of the action \eqref{eq:FirstVariationBulkLagrangian}. It has support on $\partial M$, including terms on $\B$ that typically do not vanish as $r \to \infty$ for the full range of field variations that preserve the boundary conditions. This is addressed by adding boundary terms to the action, a procedure that will be discussed in more detail for specific models later in this paper. The point we wish to make here is that the structure of the shifted presymplectic potential -- not just $\Theta^{r}$ on $\B$, but also the corner terms obtained from the components $\Theta^{a}$, which contribute at $\partial\B$ -- will determine some of the boundary terms which must be included in the action.

In the following sections we construct the codimension-2 form and identify the charges for partially on-shell fields in lower-dimensional gravitational theories. Then we apply what was learned about the structure of the presymplectic potential to obtain an action with a well defined variational principle for (Dirichlet) boundary conditions on the fields at $r \to \infty$. In each example the charges are calculated first, to emphasize that our prescription is not sensitive to boundary terms in the action, or how one carries out ``holographic renormalization'' or related procedures.

\section{2d Dilaton gravity}
\label{sec:2d}

In this section we apply our prescription for the charges to  the $UV$-family of dilaton gravity theories in 2d dimensions, for a review see \cite{Grumiller:2002nm}. The bulk Lagrangian is 
\begin{gather}\label{eq:BulkLagrangianDG}
    L_{M} = \frac{1}{2\,\kappa^{2}}\,\sqrt{-g}\,\Big(X\,R - U(X)\,(\nabla X)^{2} - 2\,V(X) \Big) ~.
\end{gather}
where $g_{\mu\nu}$ is the two-dimensional metric, $X$ is the dilaton, and $U$ and $V$ are two arbitrary functions of the dilaton.

We first present the solution space in linear dilaton Bondi gauge, and explain which equations of motion are enforced for the partially on-shell fields. The most general symmetries and transformation laws that preserve the form of the fields are determined, with special care given to an on-shell Killing vector that is present in all 2D dilaton gravity theories. When the theory is not fully on-shell, as is the case here, this diffeomorphism acts on the metric at the boundary. Incorporating it into the symmetries of the theory identifies a specific field-dependence of the symmetry parameters which naturally leads to integrable charges in the presence of corners. The adapted covariant phase space construction outlined in the previous section ensures that our results are finite as $r \to \infty$ . Carrying out this calculation before discussing the variational principle emphasizes two points. First, the procedure is independent of how ``holographic renormalization'' is implemented in the action. And second, the terms needed to cancel divergences in the codimension-2 form $k^{\mu\nu}_{\xi}$ are obtained directly from the variation of the bulk Lagrangian, at the level of the presymplectic potential -- there is no need for \emph{ad hoc} subtractions. We also note that our prescription selects a specific finite part of the codimension-2 form, and comment on its relation to the charges that were considered in \cite{Ruzziconi:2020wrb}. As pointed out in the previous section, the construction of the charges takes place without the fields being fully on-shell.

Next, we revisit the first variation of the action and discuss the variational principle in the presence of corners at spatial infinity. This leads to a discussion of holographic renormalization that generalizes the results of \cite{Grumiller:2007ju, Grumiller:2017qao} when spatial infinity is not obtained as the limit of an isocurve of the dilaton. In the case of models with no kinetic term ($U=0$) the boundary terms required for a well-defined variational principle (and a finite on-shell action) are similar to the ones derived in \cite{Ruzziconi:2020wrb}. However, the origin of these terms is more clear in the present construction.

\subsection{Solution space and symmetries in linear Bondi gauge}
\label{subsec:2dsolutions}

The first variation of the bulk Lagrangian \eqref{eq:BulkLagrangianDG} takes the form
\begin{gather}\label{eq:FirstVariation}
    \delta L_{M} = (E^{\mu\nu}\,\delta g_{\mu\nu} + E_{X}\,\delta{X})+ \partial_{\mu}\Theta^{\mu} ~,
\end{gather}
where 
the equations of motion are
\begin{align}    \label{eq:2deom}
E^{\mu\nu}& =  \frac{\sqrt{-g}}{2\kappa^{2}}\left(\nabla^{\mu}\nabla^{\nu}X - g^{\mu\nu}\,\nabla^{2}X - g^{\mu\nu}\,V + U\,\Big(\nabla^\mu X \nabla^\nu X - \frac{1}{2}g^{\mu\nu}(\nabla X)^2\Big)\right) \\
    E_{X} &= \frac{\sqrt{-g}}{2\kappa^{2}}\left( R + \partial_{X}U\,(\nabla X)^{2} +2\,U\,\nabla^{2}X - 2\partial_{X}V \right) ~.
\end{align} 
The presymplectic potential receives contributions from both the metric and the dilaton 
\begin{gather}\label{eq:DGPresymplecticPotential}
    \Theta^{\mu} = \frac{\sqrt{-g}}{2\,\kappa^{2}}\,\Big(-2U\,\nabla^{\mu}X\,\delta X + (g^{\mu\lambda}g^{\nu\sigma} - g^{\mu\nu}g^{\lambda\sigma})\big(X\,\nabla_{\nu}\delta g_{\lambda\sigma} - {\delta} g_{\lambda\sigma} \nabla_{\nu}X\big)\Big) ~.
\end{gather}

\paragraph{Linear dilaton Bondi gauge}

The coordinate gauge is partially fixed by choosing a spacelike coordinate $r$ such that the other coordinate $u$ is null, which implies $g_{rr} = 0$, and the dilaton is linear in $r$. 
\begin{align} \label{eq:BondiLineElement}
     \de s^{2} &= 2 B(u,r)\,\de u^{2} - 2\,e^{A(u,r)}\,\de u\,\de r  \\ \label{eq:FirstLinearDilaton}
    X &= r\,e^{-q(u)} + \varphi(u) ~.
\end{align}
The condition on the dilaton is the 2D equivalent of specifying the transverse metric in higher dimensional Bondi gauge. The constraint associated with fixing $g_{rr}$, which would normally be enforced by setting the $E^{rr}$ component of the metric equation of motion to zero, is relaxed. The remaining equations of motion are sufficient to completely determine the $r$-dependence of the other fields.  Solving the $E^{uu}$ and $E^{ur}$ components of the metric equations of motion, the functions $A(u,r)$ and $B(u,r)$ in the metric take the form
\begin{subequations}\label{eq:2dpartiallyonshell}
\begin{align} \label{eq:Adef}
    A(u,r) = & \,\, a(u) + Q(X) \\ \label{eq:Bdef}
    B(u,r) = & \,\, -\frac{1}{2}\,e^{Q(X)}e^{2(a(u)+q(u))}\Big(w(X) + 2\,e^{-(a(u)+q(u))}\,\partial_uX - 2\,\M(u)\Big) ~.
\end{align}
\end{subequations}
Here $Q(X)$ and $w(X)$ are familiar functions constructed from $U(X)$ and $V(X)$ as
\begin{align} \label{eq:Qw}
    Q(X)& = \int^{X} \bns dY\,U(Y) \qquad\qquad
    w(X) = -2\,\int^X dY e^{Q(Y)}\,V(Y) ~
\end{align}
The final term in $B(u,r)$ involves the Casimir\,\footnote{When the dilaton gravity admits a Poisson Sigma model formulation, $\M(u)$ is typically a Casimir function when the fields are fully on-shell.} $\M(u)$. Thus, the fields in linear dilaton Bondi gauge are described by a set of four functions $q(u)$, $\varphi(u)$, $a(u)$, and $\M(u)$. {This matches the results obtained in appendix A of \cite{Ruzziconi:2020wrb}. }

With the fields parameterized as above, the $E^{rr}$ component of the metric equation of motion becomes
\begin{gather}\label{eq:CasimirConstraint}
    E^{rr} = \frac{1}{2\kappa^{2}}\,e^{a(u)+q(u)}e^{-A(u,r)}\,\partial_u\M(u) ~.
\end{gather}
Hence, the constraint $E^{rr}=0$ associated with the choice of coordinate gauge $g_{rr} = 0$ would imply constant $\M$. As stated earlier, we relax this condition and consider the partially on-shell theory where $\M$ is instead an arbitrary function of $u$. In the static case $\pu\M=0$ the conventions above reproduce (after transforming to Schwarzschild gauge and defining a new coordinate $\tilde{r}$) the conventions of \cite{Grumiller:2007ju}.

\paragraph{Symmetries and transformation laws}

A diffeomorphism $\xi^{\mu}$ that preserves the gauge condition $g_{rr} = 0$ requires $\xi^{u}$ independent of $r$, while preserving the form of the dilaton fixes $\xi^{r}$ to also be linear in $r$,  
\begin{gather}\label{eq:Diffeo1}
    \xi = \epsilon(u)\,\partial_{u} + \Big(r\,\chi(u) + \eta(u)\Big)\,\partial_{r} ~. 
\end{gather}
Under this diffeomorphism the fields $A$ and $B$ appearing in the metric transform as 
\begin{align}
    \delta_{\xi} A &= \partial_u\epsilon+\epsilon \,\partial_uA+\eta  \partial_{r}A + \chi (1+r \partial_{r} A) \\
    \delta_{\xi}B & = 2B\partial_u\epsilon+\epsilon \, \partial_uB-e^{A} \partial_u\big(r\,\chi(u) + \eta(u)\big)+\big(r\,\chi(u) + \eta(u)\big)\partial_{r}B ~.
\end{align}
In terms of the integration functions $\epsilon$, $\chi$, and $\eta$, the set of functions of $u$ parameterizing the field configurations transform as
\begin{gather} \label{eq:DGDiffeo1}
    \delta_{\xi} q = \epsilon\,\partial_uq-\chi \qquad \delta_{\xi}\varphi = \epsilon\,\partial_u\varphi + \eta\,e^{-q} \\ \label{eq:DGDiffeo2}
    \delta_{\xi}a =\partial_u \epsilon + \epsilon\,\partial_ua + \chi \qquad   \delta_{\xi}\M = \epsilon\,\partial_{u}\M = \epsilon\,e^{-(a+q)}\,e^{A}\,2\kappa^2\,E^{rr} ~.
\end{gather}
There is one diffeomorphism of particular interest among these transformations. In two dimensions, any solution of a dilaton gravity theory possesses a Killing vector that also  leaves the dilaton invariant \cite{Schmidt:1991ws, Banks:1991mk}. For the bulk Lagrangian \eqref{eq:BulkLagrangianDG} this vector takes the form
\begin{gather}\label{eq:DGKV}
    \xi_{X} = e^{Q(X)}\,\epsilon^{\mu\nu}\,\partial_{\mu}X\,\partial_{\nu} ~,
\end{gather}
with $Q$ defined in \eqref{eq:Qw} and $\epsilon^{\mu\nu}$ related to the Levi-Civita alternating symbol by $\epsilon^{\mu\nu} = \varepsilon^{\mu\nu}/\sqrt{-g}$. By construction, this diffeomorphism leaves the dilaton invariant. However, since we are not fully on-shell, it acts non-trivially on the metric. In Bondi gauge $\sqrt{-g} = e^{A}$, so using \eqref{eq:BondiLineElement} and the fields \eqref{eq:Adef},\eqref{eq:Bdef} this vector is
\begin{gather}
    \xi_{X} = e^{-(a+q)}\partial_{u} + \Big(e^{-(a+q)}\,r\,\partial_uq - e^{-a}\,\partial_u\varphi\Big)\partial_{r} ~.
\end{gather}
This has the same general form as \eqref{eq:Diffeo1}, and from \eqref{eq:DGDiffeo1}-\eqref{eq:DGDiffeo2} we see that only $\M(u)$ transforms
\begin{gather}
    \delta_{\xi_{X}} \M = e^{-(a+q)}\,\partial_u\M ~.
\end{gather}
Requiring that $\xi_{X}$ is among the asymptotic symmetries generated by \eqref{eq:Diffeo1} suggests a specific field-dependence of the functions $\epsilon$, $\chi$, and $\eta$. If we rewrite these functions as
\begin{align}\label{eq:DGSymmetryParameterFieldDependence}
    \epsilon = & \,\, \bar{\epsilon}\,e^{-(a+q)} \,,\qquad 
    \chi =  \,\, \bar{\chi} + \bar{\epsilon}\,e^{-(a+q)}\,\partial_uq \,,\qquad 
    \eta = \,\, \bar{\eta} - \bar{\epsilon}\,e^{-a}\,\partial_u\varphi ~
\end{align}
where $\bar{\epsilon},\bar{\chi},\bar{\eta}$ are functions of $u$, then the asymptotic symmetries are generated by vectors
\begin{gather}\label{eq:AsymptoticSymmetries}
    \xi = \bar{\epsilon}\,\xi_{X} + \bar{\chi}\,r\,\partial_{r} + \bar{\eta}\,\partial_{r} ~.
\end{gather}
For instance the fields transform under this parameterization as 
\begin{gather}
    \delta_{\xi}\big(e^{a+q}\big) = \pu \bar{\epsilon} \qquad \delta_{\xi}q = -\bar{\chi} \qquad \delta_{\xi}\varphi = \bar{\eta}\,e^{-q} \qquad \delta_{\xi}\M = e^{-(a+q)}\,\bar{\epsilon}\,\pu \M ~.
\end{gather}
As we will show in the next section, there is a natural choice for the codimension-2 form $k^{ur}_{\xi}$ such that the infinitesimal charges are integrable when the symmetry parameters $\bar{\epsilon}$, $\bar{\eta}$, and $\bar{\chi}$ are taken to be field-independent.

Using the modified Lie bracket \cite{Barnich:2010eb}, the algebra of the asymptotic Killing vector is given by 
\begin{align}\label{AKV algebra}
  [ \xi_{1},\xi_{2}]_\star=\left(\bar \eta_1\,\bar \chi_2-\bar\chi_1\bar\eta_2 \right)\partial_r -\delta_1\xi_{2}+\delta_2\xi_{1}\,,
\end{align}
and for the slicing \begin{equation} \label{eq:2dintegrableslicing} \delta \bar \epsilon=\delta \bar \eta=\delta \bar \chi=0\end{equation} the algebra is three abelian Lie algebras of smooth functions with a Heisenberg central extension. Note that if we take the functions appearing in \eqref{eq:DGKV} to be state independent ($\delta\epsilon=\delta\chi=\delta \eta=0$) the algebra is the diffeomorphism along the $u$-direction in semi-direct sum with an Heisenberg algebra, see eq. (2.24) of \cite{Ruzziconi:2020wrb}.

\subsection{Covariant phase space for two dimensional dilaton gravity} 

First, let us review the construction of section \ref{sec1:covphasespacedetail} applied to dilaton gravity in two dimensions \cite{Iyer:1994ys,Ruzziconi:2020wrb,Grumiller:2021cwg}. The first variation of the bulk Lagrangian in \eqref{eq:BulkLagrangianDG} is
\begin{gather}\label{eq:FirstVariation}
    \delta L_{M} = E^{\mu\nu}\,\delta g_{\mu\nu} + E_{X}\,\delta X + \partial_{\mu}\Theta^{\mu} ~,
\end{gather}
with equations of motion given in \eqref{eq:2deom} and the presymplectic potential given by \eqref{eq:DGPresymplecticPotential}. When the fields variations are given by a diffeomorphism, integration by parts of the bulk term in \eqref{eq:FirstVariation} gives 
\begin{gather}\label{eq:DGBulkTermDiffeo1}
    E^{\mu\nu}\,\delta_{\xi} g_{\mu\nu} + E_{X}\,\delta_{\xi} X  = \xi^{\nu} N_{\nu} + \partial_{\mu}S^{\mu}_{\xi} ~.
\end{gather}
The first term on the right-hand side, which vanishes by Noether's second theorem, takes the form
\begin{gather}
    N_{\nu} = E_{X} \partial_{\nu}X + E^{\mu\lambda}\partial_{\nu} g_{\mu\lambda} - 2 \partial_{\mu} E^{\mu}{}_{\nu} 
    = 0 ~,
\end{gather}
The second term in \eqref{eq:DGBulkTermDiffeo1} is the weakly vanishing Noether current $S^{\mu}_{\xi} = \xi^{\nu} (E^{i}\,\Psi_{i})^{\mu}{}_{\nu}$, which in this case involves only
\begin{gather}
(E^{i} \Psi_{i})^{\mu}{}_{\nu} = 2\,E^{\mu\lambda}\,g_{\lambda\nu} ~.
\end{gather}
Since we have relaxed the constraint involving $E^{rr}$, this quantity and its variation have non-vanishing components which will contribute to the construction of the charges.

The presymplectic current $\omega^{\mu}$ is obtained from the total derivative term in the anti-symmetrized second variation of the bulk Lagrangian. Calculating $\omega^{\mu}$ from \eqref{eq:DGPresymplecticPotential} is straightforward, and gives
\begin{align} \nonumber
\frac{2\kappa^{2}}{\sqrt{-g}}\,\omega^{\mu} = &\,\, U\,\nabla^{\mu}X\,\delta_1g\,\delta_2 X  + 2\,U\,\delta_2 X\,\nabla^{\mu}\delta_1 X + 2\, U\,\nabla_{\nu}X\,\delta_1 X (\delta_2 g)^{\mu\nu} + \delta_1 X\,\nabla^{\mu}\delta_2 g  \\ 
& + \delta_1 g\,\nabla^{\mu}\delta_2 X + \delta_2 X\,\nabla_{\nu}(\delta_1 g)^{\mu\nu} + (\delta_2 g)^{\mu\nu}\nabla_{\nu}\delta_1 X + \frac{1}{2}\,(\delta_1 g)^{\mu\nu} \, \delta_2 g\, \nabla_{\nu} X \\ \nonumber
& + \frac{1}{2}\,X\,\Big[ \delta_2 g_{\nu\lambda} \nabla^{\mu} (\delta_1 g)^{\nu\lambda} + \delta_1 g\,\nabla^{\mu} \delta_2 g + (\delta_2 g)^{\mu\nu}\nabla_{\nu} \delta_1 g - 2 \,\delta_2 g_{\nu\lambda} \nabla^{\lambda} (\delta_1 g)^{\mu\nu} \\ \nonumber
& \qquad \qquad + \delta_2 g \nabla_{\nu} (\delta_1 g)^{\mu\nu} \Big] - \big( 1 \leftrightarrow 2 \big) ~.
\end{align}
In this expression, $\delta g = g^{\mu\nu} \delta g_{\mu\nu}$ and $(\delta g)^{\mu\nu} = g^{\mu\lambda}\,g^{\nu\sigma}\,\delta g_{\lambda\sigma} = - \delta (g^{\mu\nu})$. For the special case of a field variation $\delta_1$ that takes the same form as a diffeomorphism $\delta_{\xi}$, with $\delta_2$ arbitrary, the result \eqref{eq:OffShellSymplecticCurrentDiffeo} from the introduction implies the existence of a (not unique) codimension-2 form $k^{\mu\nu}_{\xi}$ such that
\begin{gather}\label{eq:kCondition}
\partial_{\nu}k^{\mu\nu}_{\xi} = \omega^{\mu}(\Psi;\delta_\xi\Psi,\delta\Psi) + 2\,\delta(E^{\mu\lambda} g_{\lambda\nu})\,\xi^{\nu} - \xi^{\mu}\,\big(E^{\nu\lambda}\,\delta g_{\nu\lambda} + E_{X}\,\delta X \big) ~.
\end{gather}
This result holds off-shell as explained in section \ref{sec1:covphasespace}. When the fields are fully on-shell, the presymplectic current is conserved and \eqref{eq:kCondition} reduces to $\partial_{\nu}k^{\mu\nu}_{\xi} = \omega^{\mu}(\Psi;\delta_\xi\Psi,\delta\Psi)$. This is not the case for the field configurations described in the previous section, which instead satisfy $E^{i}\delta\Psi_i=0$ through a combination of a subset of the equations of motion ($E^{uu} = E^{ur} = E_{X} = 0$) and a coordinate gauge condition ($\delta g_{rr} = 0$). In this case $\omega^{\mu}$ and $\delta(E^{\mu\lambda} g_{\lambda\nu})\xi^{\nu}$ are independently conserved. However, it is their combination appearing in \eqref{eq:kCondition} that will be relevant for the calculation of the charges.

For dilaton gravity described by \eqref{eq:BulkLagrangianDG}, we find $k^{\mu\nu}_{\xi}$ satisfying \eqref{eq:kCondition} is given by
\begin{align}\label{eq:kInitial}
 \frac{\kappa^{2}}{\sqrt{-g}}\,k^{\mu\nu}_{\xi} = & \,\, U\,\delta X\,\big(\xi^{\nu} \nabla^{\mu}X - \xi^{\mu}\,\nabla^{\nu} X\big) + \xi^{\nu}\nabla^{\mu}\delta X - \xi^{\mu}\nabla^{\nu}\delta X \\ \nonumber
    & \,\, + \frac{1}{2}\,\delta X\,\big(\nabla^{\nu}\xi^{\mu}-\nabla^{\mu}\xi^{\nu}\big) + \frac{1}{2}\big( \xi^{\mu}(\delta g)^{\lambda\nu} - (\delta g)^{\mu\lambda}\xi^{\nu}\big)\nabla_{\lambda} X \\ \nonumber
    & \,\, + \frac{1}{4}\,X\,\Big((\delta g)^{\mu\lambda}\,\big(\nabla_{\lambda}\xi^{\nu} + \nabla^{\nu}\xi_{\lambda}\big) - (\delta g)^{\lambda\nu}\,\big(\nabla_{\lambda}\xi^{\mu} + \nabla^{\mu}\xi_{\lambda}\big) \Big)~.
\end{align}
This is the same expression obtained in \cite{Ruzziconi:2020wrb} via the Barnich-Brandt procedure \cite{Barnich:2001jy , Barnich:2003xg}. Like that earlier result, this codimension-2 form is independent of the potential $V(X)$ appearing in the bulk Lagrangian. And while \eqref{eq:kInitial} includes explicit factors of the dilaton kinetic term $U(X)$, this dependence drops out when it is evaluated in linear dilaton Bondi gauge. A subtle point related to the last line of \eqref{eq:kInitial} is discussed in Appendix \ref{app:kEH}.

As explained in section \ref{sec1:covphasespace}, the charges are not obtained directly from the $ur$ component of \eqref{eq:kInitial}. It is instructive to see why this does not work. First, evaluating \eqref{eq:kInitial} for the fields \eqref{eq:BondiLineElement}-\eqref{eq:Bdef}, we find a term that is linear in $r$
\begin{gather}\label{eq:Initialkur}
    k^{ur}_{\xi} = \frac{1}{2}\,e^{-q}\,r\,\Big(\chi\,(\delta a + \delta q) + \epsilon\,(\pu a\,\delta q - \pu q\,\delta a) + \delta q\,\pu \epsilon \Big) + \ldots ~,
\end{gather}
where $\ldots$ denotes terms that are finite as $r \to \infty$. This $r$-dependence (which comes from terms in $\omega^{\mu}$ that are linear in $r$) is the obstruction to taking the $r \to \infty$ limit. A second issue, not obviously related to the first, is that the part of \eqref{eq:Initialkur} that remains finite as $r \to \infty$ is not integrable for diffeomorphisms with field dependence \eqref{eq:DGSymmetryParameterFieldDependence}. This is not a problem \textit{per se}, since a different choice of field dependence renders the finite part integrable \cite{Ruzziconi:2020wrb}. 
However, since $\xi_{X}$ is always a Killing vector when the fields are fully on-shell, one would prefer that it appears naturally among the asymptotic symmetries for the partially on-shell field configurations. As we will show in the next section, both issues are resolved by accounting for corner contributions coming from the $\omega^{u}$ component of the presymplectic current.\,\footnote{Corner contributions from components of the weakly vanishing Noether current $\xi^{\nu}\,\delta(E^{i}\,\Psi_{i})^{\mu}{}_{\nu}$ will be relevant for 3D gravity, but do not contribute here since $E^{uu} = E^{ur} = 0$.}

\subsection{Charges}

We begin this section by implementing the prescription outlined in section \ref{sec1:covphasespace}, and showing that the result is an $r$-independent codimension-2 form that leads to integrable charges. Nevertheless, there remain ambiguities in this procedure, as well as ambiguities in the slicing -- defined as the choice of field dependence of the symmetries generators -- which we examine in the second part.

The coordinate gauge condition $g_{rr} = \delta g_{rr} = 0$ leads to a $u$ component of the presymplectic potential \eqref{eq:DGPresymplecticPotential} given by
\begin{gather}\label{eq:Thetau1}
  \kappa^2\, \Theta^{u} = -\frac{1}{2}\,\delta A\,\partial_{r} X + \delta Q\,\partial_{r}X + \frac{1}{2}\,X\,\partial_{r}\delta A ~.  
\end{gather}
Since the $r$-dependence of the fields is completely determined by the equations of motion and linear condition on the dilaton, one could simply impose them and integrate the result to express $\Theta^{u}$ as a total $r$-derivative. However, we note that already in \eqref{eq:Thetau1}, integration-by-parts gives
\begin{gather}\label{eq:Thetau2}
    \kappa^2\, \Theta^{u} = \partial_{r}\Big(\frac{1}{2}\,X\,\delta A + X\,(\delta Q - \delta A) \Big) + X\,\partial_{r}\big(\delta A - \delta Q\big) ~.
\end{gather}
The result $A(u,r) = a(u) + Q(X)$, which follows from $E^{uu}=0$ and the condition that the dilaton is linear in $r$, reduces this to a total derivative $\partial_{r} Y^{ur}$ with
\begin{gather}\label{eq:FirstYur}
   \kappa^2\,  Y^{ur} = \frac{1}{2}\,\delta Q(X) \, X - \frac{1}{2}\,\delta a\,X ~.
\end{gather}
The first term is $\delta$-exact and will not contribute to $\omega^{\mu}$ or $k^{\mu\nu}_{\xi}$, though it will be important when we return to the variational problem. The second term in \eqref{eq:FirstYur} is not $\delta$-exact and will therefore impact the charges. Note that $Y^{ur}$ is defined only up to an $r$-independent piece. Here we choose to identify $Y^{ur}$ as the combination of $A$, $X$, and their variations that naturally arises from integration by parts in $\Theta^{u}$, before imposing any equations of motion. This leads to integrable charges for the slicing \eqref{eq:DGSymmetryParameterFieldDependence}, \eqref{eq:2dintegrableslicing} associated with the Killing vector $\xi_{X}$. Another choice for $Y^{ur}$, which was used in \cite{Ruzziconi:2020wrb}, keeps only the r-dependent part of \eqref{eq:FirstYur}.  
An integrable slicing also exists in that case (though it is more complicated) and yields the same charge algebra.

The prescription described in section \ref{sec1:covphasespace} is implemented at the level of the presymplectic potential by taking $\Theta^{\mu} \to \Theta^{\mu} + \partial_{\nu}\Upsilon^{\mu\nu}$, with 
\begin{gather}
   \kappa^2\, \Upsilon^{ur} = - \kappa^2\,Y^{ur} = - \frac{1}{2}\,\delta Q(X) \,X + \frac{1}{2}\,\delta a\,X ~.
\end{gather}
This shifts the components of $\Theta^{\mu}$ as 
\begin{gather}
    \Theta^{u} \to 0 \\
   \Theta^{r} \to \Theta^{r} + \frac{1}{\kappa^{2}}\,\partial_{u}\Big(\frac{1}{2}\,\delta Q(X) \,X - \frac{1}{2}\,\delta a\,X\Big) ~.
\end{gather}
The effect on the presymplectic current is to set the component $\omega^{u}$ to zero and shift the $\omega^{r}$ component by a corner term
\begin{align}\label{eq:omegarShift}
    \omega^{r} \to \omega^{r} + \frac{1}{2\kappa^{2}}\,\partial_{u}\Big(\delta_1 X\,\delta_2 a - \delta_2 X\,\delta_1 a \Big) ~.
\end{align}
This in turn changes the $ur$ component of the codimension-2 form as
\begin{gather}\label{charge ren}
    k^{ur}_{\xi} \to k^{ur}_{\xi} - \frac{1}{2\kappa^{2}}\,\Big(\delta_\xi X\,\delta a - \delta X\,\delta_\xi a \Big) ~,
\end{gather}
with the minus sign relative to \eqref{eq:omegarShift} coming from considering $k^{ur}_{\xi}$ rather than $k^{ru}_{\xi}$. The shift in $k^{ur}_{\xi}$ includes a piece that is linear in $r$, from the leading part of the dilaton, which precisely cancels the terms shown in \eqref{eq:Initialkur}. The result for $k^{ur}_{\xi}$ is now independent of $r$ and given by
\begin{align}\label{eq:AlmostThere}
  \kappa^2\, k^{ur}_{\xi} = & \,\, \epsilon(u)\,e^{a(u)+q(u)}\,\delta \M(u) + \delta\varphi(u)\,\Big(\chi(u) - \epsilon(u)\,\pu q(u)\Big) \\ \nonumber 
    & \quad + \delta q(u)\,\Big(\eta(u)\,e^{-q(u)} + \epsilon(u)\,\pu \varphi(u)\Big) ~.
\end{align}
Incorporating the field dependence \eqref{eq:DGSymmetryParameterFieldDependence} of the symmetry parameters, this reduces to
\begin{gather}\label{Charge in int slicing 1}
 \kappa^2\, k^{ur}_{\xi} = \bar{\epsilon}(u)\,\delta\M(u) + \bar{\chi}(u)\,\delta\varphi(u) + \bar{\eta}(u)\,e^{-q(u)}\,\delta q(u) ~.
\end{gather}
In the slicing \eqref{eq:2dintegrableslicing} where $\delta\bar{\epsilon}=\delta \bar{\eta}=\delta \bar{\chi}=0 $ this is integrable, $k^{ur}_{\xi} = \delta Q_{\xi}$. The charge is then
\begin{gather}\label{Charge in int slicing 2}
  Q_\xi= \frac1{\kappa^2} \left(\bar{\epsilon}(u)\,\M(u) + \bar{\chi}(u)\,\varphi(u) -\bar{\eta}(u)\,e^{-q(u)}\right) ~.
\end{gather}
Since the charges are integrable, they form a representation of the asymptotic symmetry algebra up to a possible central extension\cite{Brown:1986nw , Brown:1986ed} (see also \cite{Barnich:1991tc, Barnich:2001jy, Barnich:2007bf} for the covariant formulation of this result). Thus,
\begin{equation}
    \delta_{\xi_2}(Q_{\xi_1})=:[(Q_{\xi_1}),(Q_{\xi_2})] = Q_{\xi_{12}} ~,
\end{equation}
where ${\xi_{12}}$ is given by \eqref{AKV algebra}, and the charge algebra is given by three abelian Lie algebras with a Heisenberg central extension.

A few comments about the charges are in order. First, note that \eqref{Charge in int slicing 2} does not depend on the function $a(u)$. This is not surprising, since the fields in linear dilaton Bondi gauge involve four functions of $u$, but only three field-dependent symmetry parameters appear in the diffeomorphisms $\xi$ \eqref{eq:AsymptoticSymmetries}. 
Second, the charges are not sensitive to the specific potential $V(X)$ or kinetic term $U(X)$ appearing in the bulk Lagrangian. There is explicit dependence on $U(X)$ in the first few terms of \eqref{eq:kInitial}, but this is canceled by other terms when the full expression is evaluated in linear dilaton Bondi gauge. This is not changed by the shift in the presymplectic potential, which depends on $U(X)$ only through a $\delta$-exact term that does not affect $\omega^{r}$ or $k^{ur}_{\xi}$.

Finally, for the partially on-shell fields in linear Bondi gauge, the codimension-2 form \eqref{eq:AlmostThere} is obtained from \eqref{eq:kCondition} by evaluating $\omega^{r} + 2\,\xi^{\nu}\,\delta(E^{r\lambda}g_{\lambda\nu})$ and writing the result as a total $u$-derivative. As we saw above, when $\Theta^{u}$ is set to zero there is a corresponding change in $\omega^{r}$ which leads to the shift \eqref{charge ren}. This shift cancels the problematic term in \eqref{eq:Initialkur} linear in $r$, and also contributes a finite piece. But the contribution from $2\,\xi^{\nu}\,\delta(E^{r\lambda}g_{\lambda\nu})$ also plays an important role. It evaluates to
\begin{gather}\label{eq:deltaScurrent}
    2\,\xi^{\nu}\,\delta(E^{r\lambda}g_{\lambda\nu}) = -\frac1{\kappa^2}\,e^{a+q}\,\big(\delta\M' + \delta(a+q)\,\M'\big)\,\epsilon ~.
\end{gather}
This combines with a term in $\omega^{r}$ to produce the total derivative $\pu\big(-\epsilon\,e^{a+q}\,\delta\M\big)$, which is the first term in \eqref{eq:AlmostThere}. Without the contribution from the variation of the weakly vanishing Noether current, we would not obtain $\M(u)$ as a charge with the slicing \eqref{eq:DGSymmetryParameterFieldDependence}. In fact, an integrable charge would only be obtained in that case by requiring $\pu \M = \pu \delta \M = 0$, which is equivalent to enforcing $E^{rr} = 0$ and placing the fields fully on-shell. The contribution from the weakly vanishing Noether current is essential for capturing the full tower of charges $\M(u)$, as opposed to just the zero mode. This is important, for example, when reproducing the off-shell symmetry of the SYK dual of the Jackiw-Teitelboim model of dilaton gravity \cite{Grumiller:2017qao}.

\subsection{First variation of the action and holographic renormalization}

In the previous sections we obtained a finite and integrable codimension-2 form $k_{\xi}^{\mu\nu}$ via a shift in the presymplectic potential that set $\Theta^{u}$ to zero. Now we turn to an analysis of the variational principle, where this shift is associated with terms in the first variation of the action that have support on the corners $\partial\B$.

The goal of this section is to construct an action for general models of dilaton gravity that admits a well-defined variational problem when the fields satisfy Dirichlet boundary conditions at a spatial infinity that includes corners. We work in linear dilaton Bondi gauge, on a ``regulated'' spacetime $M$ with a boundary $\partial M$. The boundary includes a timelike component $\B$ and null components $\N_{\pm}$ that intersect $\B$ at corners $\partial \mathcal B_{\pm} = \mathcal B \cap \N_{\pm}$. This is shown schematically in the figure below. The component $\B$ is a curve of constant $r = r_{c}$ which becomes spatial infinity when the cut-off is removed by taking $r_{c} \to \infty$. Calculations involving the action and its first variation are performed on this regulated spacetime, with $r_c$ large but finite, and the $r_c \to \infty$ limit taken as the final step. 
\begin{figure}[h]
    \centering
\begin{tikzpicture}
  \draw[black] (-1,1) -- (0,2) node[pos=.5,above,sloped]{$\N_{+}$};    
  \draw[black] (-1,-1) -- (0,0)node[pos=.5,above,sloped]{$\N_{-}$};    
  \draw[black] (0,0) -- (0,2) node[pos=.5,right] {$\mathcal B$};  
\end{tikzpicture}
\caption{
}\label{}
\end{figure}

Before jumping into the construction, let us quickly review the variational principle given in \cite{Ruzziconi:2020wrb}. That analysis, tailored to dilaton gravity models with $U=0$,  accounts for corners and arrives at the action
\begin{align}\label{eq:OldActionWithCorners} \nonumber
\Gamma^{{\tiny\text{\cite{Ruzziconi:2020wrb}}}} = & \,\, \frac{1}{2\kappa^{2}}\int_{M} \nts d^{2}x \sqrt{-g}\,\Big(X\,R - 2\,V(X)\Big)\\& + \frac{1}{\kappa^{2}}\,\int_{\B}d^{1}x \sqrt{-\gamma} \big( X\, K+\mathcal L_w+ \mathcal L_c+ \mathcal L_n \big) ~.
\end{align}
Here $K$ is the extrinsic curvature of a constant $r$ curve embedded in $M$ with outward pointing unit normal $n^{\mu}$, and 
\begin{align}\mathcal  L_w = -\sqrt{w+2e^{-(a + q)}\pu X },  \quad  \mathcal L_c = D_{u}(v^{u}\, X),\quad \mathcal L_n= - \sqrt{\gamma} X  v^\mu a_\mu ~.
\end{align} 
In the last two functions, $v$ is a unit vector field ($v^a v_a = \gamma_{uu} v^u v^u = -1$) tangent to {$r$\,=\,constant} leaves, $D$ is the covariant derivative compatible with the induced metric $\gamma_{uu}$ on $\B$, and $a_\mu$ is the ``acceleration'' $n^\nu\nabla_\nu n_\mu$. This action is finite on-shell, has a well-defined flat limit, and is stationary for Dirichlet boundary conditions. But since it assumes $U=0$ it is not suitable for the full range of dilaton gravity models we consider. Furthermore, the motivation for some of the boundary terms is \emph{ad hoc}, so even though a number of generalizations to $U \neq 0$ suggest themselves, it is not immediately clear if they are all on equal footing. Rather than starting from this action and trying to modify it, we will instead use variational arguments and holographic renormalization techniques to build an appropriate action from the ground up, paying careful attention to contributions at the corners and on components of $\partial M$ other than $\B$. The resulting action will reduce (in the $r_c \to \infty$ limit) to \eqref{eq:OldActionWithCorners} for models with $U = 0$.

For our purposes, a theory described by an action $\Gamma$ and boundary conditions on the fields has a well-defined variational principle if field configurations satisfying $E^{i}\,\delta\Psi_{i} = 0$ are local extrema of $\Gamma$ among all nearby field configurations with the same boundary conditions. An essential point is that $\delta \Gamma = 0$ for \textit{any} field variation consistent with the boundary conditions, and not just field variations that satisfy additional and more restrictive conditions.\,\footnote{This is a common problem when constructing an action for fields that satisfy conditions at spatial infinity. An example is four-dimensional gravity with asymptotically flat boundary conditions. A meaningful variational principle should distinguish the Schwarzschild solution from competing field configurations with similar asymptotic fall-off; i.e., metrics whose $g_{tt}$ components differ from Schwarzschild at $\O(1/r)$ as $ r\to \infty$. But the variation of the Einstein-Hilbert action plus the Gibbons-Hawking-York boundary term does not vanish unless the metric variation $\delta g_{tt}$ falls off \textit{faster} than $1/r$ \cite{Regge:1974zd}. To address this problem, one must add additional boundary terms to the action \cite{ Mann:2005yr, Mann:2006bd, Mann:2008ay}. }
For the variational principle to have any meaning, it must be able to distinguish solutions of the equations of motion from all competing field configurations with similar asymptotic behavior. This means that field variations with any asymptotic behavior that preserves the boundary conditions must be allowed.

The focus here will be on Dirichlet boundary conditions for the fields at $\B$. In that case $\delta \Gamma$ evaluated on {$E^{i}\,\delta\Psi_{i} = 0$} should be a functional of the fields $\Psi_{i}$ and their variations $\delta\Psi_{i}$ on $\B$, as well as derivatives $\pu\delta\Psi_i$ of the field variations tangent to $\B$. But it should not depend on the derivatives $\partial_{r}\delta\Psi_{i}$ of the field variations in the direction normal to $\B$. Working in linear dilaton Bondi gauge, with fields $\Psi_{i} = A$, $B$, and $X$, we take Dirichlet boundary conditions to fix the boundary data $a(u)$, $q(u)$, and $\varphi(u)$ in \eqref{eq:FirstLinearDilaton}-\eqref{eq:2dpartiallyonshell}. Thus, when $E^{i}\delta \Psi_{i} = 0$, a well-defined variational principle should have $\delta \Gamma = 0$ for all field variations that preserve $g_{rr} = 0$ and do not change this boundary data. An immediate and useful consequence is that changing the boundary data should produce a finite  (as $r_c \to \infty$) change in the action.\,\footnote{If a change in the boundary data caused $\delta\Gamma$ to diverge as $r_c \to \infty$, then a suitably rescaled field variation would give a finite and non-zero $\delta \Gamma$, while also falling off fast enough to preserve the boundary data. This would contradict the defining property of the variational principle, that $\delta\Gamma = 0$ for any field variation consistent with the boundary conditions.} It will be convenient in the following section to use this fact when analyzing the properties of the action, since changes in the boundary data $\delta a$, $\delta q$, and $\delta \varphi$ producing a finite response $\delta \Gamma$ is a necessary (though not sufficient) condition for a well-defined variational principle.

It is easy to see that an action based only on the bulk Lagrangian \eqref{eq:BulkLagrangianDG} does not lead to a well-defined variational principle. The variation of the bulk Lagrangian, evaluated on $E^{i}\delta\Psi_{i} = 0$, is just the total derivative of the presymplectic potential. Integrated over $M$, it gives terms on $\B$ (and $\partial\B$) that include factors which diverge as $r_c \to \infty$. As a result, the variation of the bulk action on its own does not vanish for all field variations preserving the boundary data. For example, in the Jackiw-Teitelboim model \cite{Jackiw:1984, Teitelboim:1984} ($U=0$, $V=-X$) there are terms in $\Theta^{r}$ proportional to $r^{2}\,\delta A$ and $r\,\delta A$. These vanish only if $\delta A$ falls off faster than $1/r^{2}$, which is much more restrictive than the condition $\delta a = 0$. This can be addressed by supplementing the action with boundary contributions whose variations cancel problematic terms on $\partial M$ coming from $\partial_{\mu}\Theta^{\mu}$ in the bulk. As long as the problematic terms on $\partial M$ are $\delta$-exact (total variations of some combination of the fields) it will be possible to cancel them against the variation $\delta L_{\partial M}$ of appropriate boundary terms.

In the following, we restrict our attention to models with potentials $U(X)$ and $V(X)$ such that $w(X) \to \infty$ as $X \to \infty$. This includes, for example, models with AdS$_2$ and Minkowski asymptotics. We also assume that $\partial_{u} X / w \sim X / w \to 0$ as $X \to \infty$. Both conditions can be relaxed, but doing so would require a qualitatively different construction than the one we present here. Additional details, and comments on the regime of applicability of some of our results, can be found a the end of appendix \ref{app:2dhamiltonjacobi}.

\subsubsection*{Corner contribution from the bulk symplectic potential}

A complication related to this last point arises when $\partial M$ includes corners. Consider the $\partial_{r}\Theta^{r}$ part of $\delta L_{M}$. When integrated over $M$ it produces terms on $\B$. In Bondi gauge they are 
\begin{align}\label{eq:InterimThetar}
  \kappa^2\, \Theta^{r} = &\,\, \delta\big(X\,e^{-A}\,\partial_{r}B\big) + \big(2\,e^{-A}\,B\,U\,\partial_{r}X - e^{-A}\,\partial_{r}B \big)\,\delta X -e^{-A}\,\partial_{r}X\,\delta B \\ \nonumber
    & \quad + \partial_uX\,\delta\big(Q - A \big) + \partial_{u}\Big(\frac{1}{2}\,X\,\delta A\Big) ~.
\end{align}
The first term includes a $\partial_{r}\delta B$ contribution, which is not consistent with Dirichlet boundary conditions, while the other terms do not vanish for all field variations that preserve the boundary conditions due to various factors that diverge as $r_c \to \infty$. For fields of the form \eqref{eq:FirstLinearDilaton}-\eqref{eq:2dpartiallyonshell} these problematic terms reduce to $\delta$-exact expressions that will be addressed with appropriate boundary terms. However, the final term in \eqref{eq:InterimThetar} -- a total $u$-derivative with support on $\partial\B$ -- includes an $r_c \to \infty$ divergence proportional to $X\,\delta a$. This corner term is \emph{not} $\delta$-exact and therefore cannot be canceled by adding a boundary term to the action. Instead, the resolution comes from the $\Theta^{u}$ component of the presymplectic potential. The integral over $M$ of $\partial_{u}\Theta^{u}$ gives terms on $\N_{\pm}$. As we saw previously, $\Theta^{u}$ in Bondi gauge is a total $r$-derivative. When integrated over $\N$ it gives contributions at the corners that can be written as $\partial_{u}Y^{ur}$ on $\B$ with
\begin{align}
       \kappa^2\, Y^{ur} = &\,\, \frac{1}{2}\,X\,\delta A + X\,\delta(Q-A) = X\,\delta Q - \frac{1}{2}\,X\,\delta A  
\end{align}
Thus, once the full total derivative $\partial_{\mu}\Theta^{\mu}$ in the variation of the bulk Lagrangian is accounted for, the terms on $\B$ are
\begin{align}\label{eq:AllBulkTermsOnB}
   \kappa^2\,  ( \Theta^{r} + \partial_{u} Y^{ur}) = &\,\, \delta\big(X\,e^{-A}\,\partial_{r}B\big) + \big(2\,e^{-A}\,B\,U\,\partial_{r}X - e^{-A}\,\partial_{r}B \big)\,\delta X \\ \nonumber
    & \quad -e^{-A}\,\partial_{r}X\,\delta B +\partial_u X\,\delta\big(Q - A \big) + \partial_{u}\big(X\,\delta Q\big) ~.
\end{align}
The parts of this expression which cause problems in the $r_c \to \infty$ limit are $\delta$-exact and will now be addressed with appropriate boundary terms.

\subsubsection*{Gibbons-Hawking-York term}

The first term in \eqref{eq:AllBulkTermsOnB} is incompatible with Dirichlet boundary conditions because it includes a contribution proportional to $\partial_{r}\delta B$. This is remedied by adding the dilaton gravity version of the Gibbons-Hawking-York (GHY) term \cite{York:1972sj, Gibbons:1977ue} on $\B$. 
\begin{gather}
\kappa^2\, L_\text{GHY} = \sqrt{-\gamma}\,X\,K = - X\,e^{-A}\,\partial_{r}B - X\,\Big(\partial_u A-\frac{\partial_uB}{2B}\Big)~.
\end{gather}
The dependence on the normal derivative of the metric eliminates the $\partial_{r}\delta B$ term in \eqref{eq:AllBulkTermsOnB}. Combining all terms with support on $\B$ up to this point, we have
\begin{align}
   \kappa^2\,( \Theta^{r} + \partial_{u} Y^{ur} + \delta L_{GHY}) = & \,\, \big(2\,e^{-A}\,B\,U\,\partial_{r}X - e^{-A}\,\partial_{r}B \big)\,\delta X -e^{-A}\,\partial_{r}X\,\delta B \\ \nonumber
    & \quad + \partial_u X\,\delta\big(Q-A\big) - \delta\Big(X\,\Big(\partial_uA-\partial_uQ-\frac{\partial_uB}{2B}\Big)\Big) ~,
\end{align}
 Evaluating this for the solution \eqref{eq:2dpartiallyonshell} and using $\pu\big(X\,\delta Q(X)\big) = \delta\big(X\,\pu Q(X)\big)$
\begin{align}\label{eq:TermsOnB1}
 \kappa^2\,(   \Theta^{r} + \partial_{u} Y^{ur} + \delta L_{GHY} )= & \,\, \delta\big(e^{a+q}\,w\big) + \partial_{u}\big(\delta X\big) -\delta\Big(X\,\partial_{u}\log\Big(\frac{e^{A-Q}}{\sqrt{-2\,B}}\Big)\Big) \\ \nonumber
    & \quad + \partial_u \varphi\,\delta q - \partial_uq\,\delta\varphi -e^{-(a+q)}\,\delta\big(e^{2(a+q)}\,\M\big) 
\end{align}
The terms in the first line on the RHS diverge as $r_c \to \infty$ for variations of the boundary data, and hence do not vanish for all field variations consistent with our boundary conditions. However, these terms are all $\delta$-exact and can be addressed by additional boundary contributions to the action. The terms in the second line are zero for variations that preserve the boundary conditions, \textit{except} for a non-zero term proportional to $\delta \M$. 

\subsubsection*{Boundary terms in the presence of corners}

To understand the remaining boundary terms, let us now write an action with the form suggested by \cite{Lehner:2016vdi} for spacetimes with a boundary that includes both timelike and null parts, generalized to dilaton gravity in two dimensions. It consists of a bulk integral over $M$ as well as contributions on the boundary components $\N_{\pm}$ and $\B$.
\begin{empheq}{align}\label{eq:ActionWithCorners}
    \Gamma = & \,\, \frac{1}{2\kappa^{2}}\int_{M} \nts d^{2}x \sqrt{-g}\,\Big(X\,R - U(X)\,\big(\nabla X \big)^{2} - 2\,V(X)\Big) + \frac{1}{\kappa^{2}}\,\int_{\B}d^{1}x \sqrt{-\gamma}\,X\,K \\ \nonumber
    & \qquad + \frac{1}{\kappa^{2}}\int_{\N_{\pm}} d\lambda\,X\,\mathcal{K} + \frac{1}{\kappa^{2}}\,\int_{\B}d^{1}x \,\Big(L_{X\partial\B} + L_{\text{{CT}}}\Big) ~.
\end{empheq}
As before, $\gamma$ is the induced metric on $\B$ and $K = \nabla_{\mu}n^{\mu}$ is the extrinsic curvature formed from the outward-pointing spacelike unit vector $n^{\mu}$. The integrals on the null components of the boundary involve a parameter $\lambda$ along a null generator $k_{\mu}$, while $\mathcal{K}$ measures the failure of this parameter to be affine according to
\begin{gather}
    k^{\nu}\nabla_{\nu}k^{\mu} = \K\,k^{\mu} ~,
\end{gather}
or $\mathcal{K} = \nabla_{\mu}k^{\mu}$. The last set of terms in \eqref{eq:ActionWithCorners} are a boundary term $L_{X\partial\B}$ that is needed when $\B$ has corners, along with a ``boundary counterterm'' $L_\text{CT}$ that implements a holographic renormalization prescription explained in the next section. Note that $L_{X\partial\B}$ is not necessarily a pure corner term. Rather, it represents terms which may be needed (in addition to the other boundary terms) when there is a corner $\B \cap \N$.

An ambiguity in the action \eqref{eq:ActionWithCorners} is that it depends on the definition of the null generator $k_{\mu}$, which is not unique. We fix part of this ambiguity as in \cite{Lehner:2016vdi} by demanding affine $\lambda$, so that $\mathcal{K} = 0$ and the integrals over $\N_{\pm}$ vanish. Then motivated by the construction in \cite{Lehner:2016vdi}, the term $L_{X\partial\B}$ on $\B$ takes the form
\begin{gather}
    \kappa^2\,L_{X\partial\B} = \sqrt{-\gamma}\,X\,v^{u}\partial_{u}\log\big(n^{\mu}\,k_{\mu}\big) ~,
\end{gather}
where $v^{u} = 1/\sqrt{-\gamma}$ is the timelike unit vector tangent to $\B$. Requiring affine $\lambda$ gives $k_{\mu} = -\beta(u)\,\delta_{\mu}{}^{u}$ with arbitrary $\beta(u)$. Contracting this with the unit vector $n^{\mu}$ normal to a curve of constant $r$, we have
\begin{gather}
    \kappa^2\,L_{X\partial\B} = X\,\pu \log\Big(\frac{\beta(u)}{\sqrt{-2B}}\Big) ~.
\end{gather}
The remaining ambiguity in the choice of $k_{\mu}$ is fixed by comparing this boundary term with the third term on the first line of \eqref{eq:TermsOnB1}. An $r_c \to \infty$ divergence cancels if $\beta(u) = e^{A-Q}$, which fixes (up to an irrelevant constant factor) the choice of null generator 
\begin{gather}\label{eq:nullnormaldef}
    k_{\mu} = -e^{A(u,r)-Q(X)}\,\delta_{\mu}{}^{u} ~.
\end{gather}
The final form of $L_{X\partial\B}$ is then
\begin{gather}\label{eq:Lcorner}
    \kappa^2\,L_{X\partial\B} = \sqrt{-\gamma} X\,v^{u}\,\partial_{u}\log\big(n^{\mu}\,k_{\mu}\big) = X\,\partial_{u}\log\left(\frac{e^{A(u,r)-Q(X)}}{\sqrt{-2B(u,r)}}\right) ~.
\end{gather}
Schematically, this can be thought of as the dimensional reduction of a higher dimensional corner term, with the dilaton identified with the transverse part of the metric. 

\subsubsection*{Boundary counterterms for dilaton gravity models}

The derivation of the boundary counterterms $L_\text{CT}$ is described in Appendix \ref{app:2dhamiltonjacobi}. Following the Hamilton-Jacobi approach used in \cite{Grumiller:2007ju}, the counterterm Lagrangian $L_{\text{CT}}$ is obtained as a solution of the radial Hamiltonian constraint in a derivative expansion on $\B$. There are no curvature invariants built from $\gamma$ and its derivatives along $\B$, so this amounts to an expansion in derivatives of $X$ which takes the form
\begin{gather}\label{eq:CT0and2derivatives}
    L_{\text{{CT}}}(\gamma,X) = \frac1{\kappa^2} \sqrt{-\gamma}\,\Big(F_{0}(X) + F_{2}(X)\,\gamma^{uu}(\partial_{u}X)^{2} + \ldots \Big) ~.
\end{gather}
The number of terms in this expansion needed for the renormalization of a particular model will depend on the $r \to \infty$ behavior of the functions $Q(X)$ and $w(X)$ constructed from $U(X)$ and $V(X)$. The derivation in Appendix \ref{app:2dhamiltonjacobi} gives the first two terms in the expansion as
\begin{gather}\label{eq:HJCT}
  \kappa^2\,L_{\text{{CT}}}(\gamma,X) =- \sqrt{-\gamma}\,\Big(\sqrt{e^{-Q(X)}w(X)} + \frac{1}{2\sqrt{e^{-Q(X)}w(X)}}\,D^{u}X\,D_{u}X \Big) + \ldots ~,
\end{gather}
where $D$ is the derivative along $\B$ compatible with the induced metric $\gamma$. In the analysis below, we will focus on the wide class of models that require only these first two terms; i.e., models for which the subsequent terms in the boundary derivative expansion of $L_{\text{CT}}$ approach zero as $r_c \to \infty$. It should be emphasized that, while the boundary counterterms are normally presented as a way of systematically addressing $r_c \to \infty$ problems with the variational principle, $L_\text{CT}$ also makes a finite contribution to both the action and its first variation. Evaluating \eqref{eq:HJCT} for the partially on-shell fields, the relevant terms are
\begin{gather}
 \kappa^2   L_\text{CT} \sim -w(X)e^{a+q} - \partial_{u}X + e^{a+q}\M + \frac{(\partial_{u}X)^{2}}{e^{a+q}w(X)} + \ldots ~.
\end{gather}
The roles played by finite parts of this expression are explained below.

\subsubsection*{Summary: Renormalized action and first variation of the action}
To summarize, the following action gives a well-defined variational principle for general dilaton gravity models with fields satisfying Dirichlet boundary conditions at a spatial infinity with corners
\begin{empheq}[box=\fbox]{align}
\label{eq:ActionWithCornersFinal}
    \Gamma = & \,\, \frac{1}{2\kappa^{2}}\int_{M} \nts d^{2}x \sqrt{-g}\,\Big(X\,R - U(X)\,\big(\nabla X \big)^{2} - 2\,V(X)\Big) + \frac{1}{\kappa^{2}}\,\int_{\B}d^{1}x \sqrt{-\gamma}\,X\,K \\ \nonumber
    & \qquad + \frac{1}{\kappa^{2}}\,\int_{\B}d^{1}x \,\Big(L_{X\partial\B} + L_{\text{{CT}}} \Big) ~.
\end{empheq}
The boundary term $L_{X\partial\B}$ is given in \eqref{eq:Lcorner}, and $L_{\text{{CT}}}$ is given in \eqref{eq:HJCT}.

Using the results of the previous sections to evaluate the first variation of \eqref{eq:ActionWithCornersFinal} in linear dilaton Bondi gauge, the terms with support on $\B$ are
\begin{gather}\label{eq:FinalDeltaGamma}
    \delta \Gamma \Big|_{\B} = \frac{1}{\kappa^{2}}\,\int_{\B}\nts du\,\left( \delta q\,\partial_{u}\varphi - \delta\varphi\,\partial_{u}q - \delta(a+q)\,e^{a+q}\,\M + \delta\Big(e^{-(a+q)} \frac{(\partial_{u}X)^2}{w}\Big) \right) ~.
\end{gather}
With our assumptions about the falloff of $w$ this vanishes for variations that preserve the Dirichlet boundary data ($a$, $q$, and $\varphi$), and it is finite for variations of the boundary data. A term in \eqref{eq:TermsOnB1} proportional to $\delta \M$ has canceled against a similar finite term coming from $L_\text{CT}$. This is important, since $\M$ is not part of the boundary data and hence $\delta\Gamma$ must vanish for a variation $\delta\M$. Note that the last term in \eqref{eq:FinalDeltaGamma}, which is the only part of the expression that explicitly depends on (integrals of) the potentials $V(X)$ and $U(X)$, is delta-exact and hence does not contribute to the symplectic current. In the case of Jackiw-Teitelboim gravity and related models, this term is needed to obtain the expected Schwarzian action for the dual theory \cite{Grumiller:2017qao}.

Finally we can also show that the limit $r_c \to \infty$ does not produce any divergences in the (partially) on-shell action. This gives
\,\footnote{In \cite{Ruzziconi:2020wrb}, there was a finite term $-e^{a+q}\M$ missing in the computation of the on-shell action.}
\begin{gather}\label{eq:2donshellaction}
    \Gamma = \frac{1}{\kappa^{2}}\,\int_{\B}\nts du\,\left( -e^{a+q}\,\M + e^{-(a+q)}\,\frac{(\partial_{u}X)^2}{w} \right) + \ldots ~,
\end{gather}
where ``$\ldots$'' refers to contributions to the action from integration-by-parts terms evaluated in the interior of $M$; i.e. total $r$ derivatives that may also give a finite contribution at $r=0$, the horizon of a black hole, or similar. When considering $\delta \Gamma$ we are only interested in terms with support on $\B$, since the fields satisfy boundary conditions at spatial infinity. For applications that make use of the on-shell action, the ``$\ldots$'' terms in \eqref{eq:2donshellaction} are just as important as the contributions on $\B$. Those terms cannot be evaluated without having a specific model and solution in mind, as they depend on what happens in the interior of $M$. The point here is simply that there are no terms in the action that diverge as $r_c \to \infty$.

\section{3d Einstein gravity}
\label{sec:3d}

In this section we revisit the results of \cite{Geiller:2021vpg} where a new gauge for three dimensional Einstein gravity, dubbed the Bondi-Weyl gauge, was introduced.\footnote{One can show that this is precisely the three dimensional analog of the partial Bondi gauge introduced later in four dimensions \cite{Geiller:2022vto}.} This is an enhancement of the Bondi gauge where the condition on the determinant of the transverse metric is relaxed. 
The space of partially on-shell field configurations in Bondi-Weyl gauge is characterized by four charges, which is two more than in the usual Bondi gauge \cite{Ruzziconi:2020wrb}. Previous efforts to calculate the charges in this gauge encountered two problems: the ``obvious'' choice of codimension-2 form diverges as $r\to \infty$, and its finite part is not integrable. These issues were addressed using the intrinsic corner ambiguities of the covariant phase space \emph{\`a la} Iyer-Wald, along with a change in the field dependence of the symmetries generators. We show in this section that this is not necessary when corner contributions in the symplectic potential are taken into account, as described in section \ref{sec1:covphasespace}. Moreover, the finite corner term introduced in \cite{Geiller:2022vto} to restore integrability appears naturally in our construction, in the same way as the dilaton gravity example of section \ref{sec:2d}. 

We begin by reviewing the solution space of the Bondi-Weyl gauge. This is followed by a quick review of the covariant phase space in Einstein gravity. Applying the results of section \ref{sec1:covphasespace} we recover the same charges as in \cite{Geiller:2021vpg}. Finally, we present a new variational principle for three-dimensional gravity with a negative cosmological that accounts for corners, and also admits a straightforward flat-space limit. 

The discussion in this section is carried out in Bondi-Weyl gauge, but many of our results hold for other choices of gauge. In appendix \ref{app:3dFG} we consider the variational properties of the action in the more familiar Fefferman--Graham gauge \cite{fefferman1985conformal}, and show that corner contributions from the symplectic potential cancel divergent total derivative terms on the boundary which cannot be discarded in the presence of corners.

\subsection{Solution space and symmetries in Bondi-Weyl gauge}
\label{subsec:3dsolutions}

We consider Einstein gravity in three dimensions with a negative cosmological constant. The bulk Lagrangian is 
\begin{equation}\label{eq:3dBulkLagrangian}
  L_M =  \frac1{2\kappa^{2}} \sqrt{-g}\,\left(R+ \frac{2}{\ell^2} \right) \,,
\end{equation}
where $\ell$ is the AdS radius.\,\footnote{The results of this section assume a negative cosmological constant, and also hold in the flat limit $\ell \to \infty$. A similar construction can be carried out in the case of a positive cosmological constant.}
Its variation \begin{gather}\label{eq:3dFirstVariation}
    \delta L_{M} = E^{\mu\nu}\,\delta g_{\mu\nu} + \partial_{\mu}\Theta^{\mu} ~,
\end{gather}
gives the equations of motions  
\begin{equation}\label{eq:3deom}
  E^{\mu\nu}:=  \frac{\delta L_M}{\delta g_{\mu\nu}}=-\frac{\sqrt{-g}}{2\kappa^2} \left(R^{\mu\nu} - \frac{1}{2}\,g^{\mu\nu} R - \frac1{\ell^2}g^{\mu\nu}\right)\,,
\end{equation}
and the symplectic potential 
\begin{gather}\label{eq:3dPresymplecticPotential}
\Theta^\mu=\frac{\sqrt{-g}}{2\kappa^2} \big(\nabla_\nu(\delta g)^{\mu\nu}-\nabla^\mu(\delta g)^{\nu}{}_{\nu}\big)\,.
\end{gather}

\paragraph{Bondi-Weyl gauge} 
The line element in the Bondi-Weyl gauge is
\begin{gather}\label{eq:3dlineelement}
\de s^2 = -2e^{2\beta(u,\phi)}\,\de u\,\de r + \mathcal V(u,r,\phi) \, e^{2\beta(u,\phi)}\,\de u^2 + {\mathcal W}^2(u,r,\phi) \Big(\de\phi-\U(u,r,\phi)\,\de u\Big)^2\,,
\end{gather}
with $g_{rr} = g_{r\phi} = \partial_{r} g_{ur} = 0$ imposed as gauge conditions. The condition $\partial_{r} g_{ur} = 0$ can be relaxed, but this will not enlarge the solution space and many of the calculations that follow are simpler with this condition in place. The coordinate $\phi$ in \eqref{eq:3dlineelement} is taken to be periodic: $\phi \sim \phi + 2\pi$. Later on we will use this to discard total $\phi$-derivatives appearing in integrals, under the assumption that the fields are single-valued. Solving the components of the equations of motion associated with the non-zero components of the metric gives\footnote{The notation here is different than in \cite{Geiller:2021vpg}, with $C=H^{\text{\cite{Geiller:2021vpg}}}$, $\mathcal V=V^{\text{\cite{Geiller:2021vpg}}}/r$, and $\mathcal W^2=r^2 \mathcal W^{\text{\cite{Geiller:2021vpg}}}$. The functions $M,N$ have been shifted as \begin{equation*}
N = N^{\text{\cite{Geiller:2021vpg}}} +2\,C\,\beta'\,,\qquad M= M^{\text{\cite{Geiller:2021vpg}}} + \frac{e^{2\beta}C^{2}}{2\ell^2} - C\,\big(U'+U\varphi'+\partial_u\varphi\big) ~.
\end{equation*}} 
\bsub
\be
E^{uu}=0\ &\Rightarrow\ {\mathcal W}= e^{\varphi}\left(r-C\right),\label{W solution}\\
E^{u\phi}=0\ 
&\Rightarrow\ \U= U+\frac{e^{2(\beta-\varphi)}}{(r-C)}\,\left( 2\,\beta'-\frac{N}{(r-C)}\right) \label{U solution}\\
E^{\phi\phi}= E^{ur} = 0\ 
\nonumber
&\Rightarrow\ \mathcal V= -\frac{e^{2\beta}}{\ell^2}(r-C)^2 -2 \,(U'+U\varphi'+\partial_u\varphi)(r-C) \\&\phantom{\ \Rightarrow\ V=}+2 \,\big(M + 2\,e^{2(\beta-\varphi)}(\beta')^2\big)-e^{2(\beta-\varphi)}\left(2\,\beta'-\frac{N}{(r-C)}\right)^{2} \label{V solution}
\ee
\esub
where $C,U,\beta,\jf$ are functions of $(u,\phi)$, and a prime $'$ denotes a partial derivative with respect to $\phi$. The two remaining components, $E^{rr}$ and $E^{r\phi}$, would typically be enforced as constraints on the functions $M$ and $N$, associated with the choice of coordinate gauge. However, we do not impose these constraints. This leaves a total of six arbitrary functions of $(u,\phi)$ parameterizing the partially on-shell fields in Bondi-Weyl gauge.

\paragraph{Symmetries and transformation laws} 
The asymptotic Killing vectors $\xi^{\mu}$ of the metric \eqref{eq:3dlineelement} take the form
\bsub\label{AKV}
\be
\xi^u&=f,\\
\xi^\phi&=g-\f{e^{2(\beta-\varphi)}}{r-C}f',\\ \nonumber
\xi^r&=(r-C)(h+Uf'-g'-g\varphi'-f\partial_u\varphi)-k
\\&\phantom{=\ }
+e^{2(\beta-\varphi)}\left(f''+f'(4\beta-\varphi)'-\f{N}{r-C}f'\right).
\ee
\esub
They depend on four arbitrary functions of $(u,\phi)$ -- denoted $f$, $g$, $h$ and $k$ -- which may be field-dependent. Under these diffeomorphisms, the fields transform as 
\bsub\label{variations}
\be
\delta_\xi\beta&=f\,\partial_u\beta+g\,\beta' + U f'+\f{1}{2}\big(\partial_uf-f\partial_u\varphi-g'-g\varphi'+h\big),\\
\delta_\xi\varphi&=h,\\
\delta_\xi U&=f\,\partial_u U + g \,U'+U \big(\partial_uf-g'+Uf' \big)-\partial_ug+\f{e^{2(2\beta-\varphi)}}{\ell^2}f',\\
\delta_\xi C&=f\,\partial_u C + g \,C'+k.
\ee
\esub
The transformations $\delta_\xi M$ and $\delta_\xi N$ are more involved, but can be easily derived.

Using the modified Lie bracket \cite{Barnich:2010eb} these vector fields satisfy the commutation relations
\be\label{modified bracket}
\big[\xi(f_1,g_1,h_1,k_1),\xi(f_2,g_2,h_2,k_2)\big]_\star
&=\big[\xi(f_1,g_1,h_1,k_1),\xi(f_2,g_2,h_2,k_2)\big]-\delta_{\xi_1}\xi_2+\delta_{\xi_2}\xi_1\cr
&=\xi(f_{12},g_{12},h_{12},k_{12}),
\ee
where
\bsub\label{commutation relations 3d}
\be
f_{12}&=f_1\partial_uf_2+g_1f_2'-\delta_{\xi_1}f_2-(1\leftrightarrow2),\\
g_{12}&=f_1\partial_ug_2+g_1g'_2-\delta_{\xi_1}g_2-(1\leftrightarrow2),\\
h_{12}&=-\delta_{\xi_1}h_2-(1\leftrightarrow2),\\
k_{12}&=f_1\partial_uk_2+g_1k_2'-\delta_{\xi_1}k_2-(1\leftrightarrow2).
\ee
\esub
When the functions parameterizing the diffeomorphism \eqref{AKV} are taken to be independent of the fields, $\delta f=\delta g=\delta h=\delta k=0$, the algebra is $\left( \text{Diff}(C^2)^{(f,g)} \loplus (\C^\infty)^{(k)}(C^2)\right) \oplus(\C^\infty)^{(h)}(C^2)$, where $C^2$ is the cylinder spanned by $(u,\phi)$ and $(\C^\infty)(C^2)$ denotes the smooth functions over $C^2$.

\subsection{Covariant phase space for three dimensional gravity}
We consider the covariant phase space, reviewed in section \ref{sec1:covphasespacedetail}, for Einstein gravity in three dimensions.\,\footnote{In fact, the expressions in this subsection are valid for pure gravity in any spacetime dimension.}

The presymplectic potential for the Einstein-Hilbert action was given in \eqref{eq:3dPresymplecticPotential}. The antisymmetrized second variation of the action then gives the presymplectic current
\begin{align}
    \omega^{\mu}(g; \delta_1 g, \delta_2 g) = \frac{\sqrt{-g}}{2\kappa^{2}}\,\left[ \vphantom{\frac{1}{2}}\right. & \,\,\frac{1}{2}\,g^{\mu\lambda}\Big( g^{\alpha\beta} g^{\sigma\nu} - g^{\alpha\sigma} g^{\beta\nu} \Big) + \frac{1}{2}\,g^{\beta\nu}\Big( g^{\alpha\lambda} g^{\sigma\mu} - g^{\alpha\sigma} g^{\lambda\mu} \Big) \\ \nonumber& \,\, + \left. \frac{1}{2}\,g^{\lambda\nu}\Big( g^{\mu\alpha} g^{\beta\sigma} - g^{\alpha\beta} g^{\sigma\mu} \Big) \right]\,\Big(\delta_2 g_{\alpha\beta}\,\nabla_{\sigma} \delta_{1} g_{\lambda\nu} - ( 1 \leftrightarrow 2 ) \Big) ~.
\end{align}
The other ingredient needed for the construction of the codimension-2 form is the weakly vanishing Noether current. Contracting the bulk term in the variation of the Lagrangian \eqref{eq:3dBulkLagrangian} with a diffeomorphism yields
\begin{gather}
    E^{\mu\nu}\,\delta_{\xi} g_{\mu\nu}= \xi^{\nu} N_{\nu} + \partial_{\mu}S^{\mu}_{\xi} ~,
\end{gather}
which gives the Noether identities and the weakly vanishing Noether current as 
\begin{align}
   N_{\nu} & = - 2\,\partial_{\mu} E^{\mu}{}_{\nu} + E^{\mu\lambda}\,\partial_{\nu} g_{\mu\lambda}= 0 ~\,,\qquad
   S_{\xi}^\mu= 2 \,E^{\mu\lambda}\,g_{\lambda\nu}\, \xi^\nu ~.
\end{align}
Then \eqref{eq:OffShellSymplecticCurrentDiffeo} implies the existence of a non-unique codimension-2 form $k^{\mu\nu}_{\xi}$ such that
\begin{gather}\label{eq:3dkCondition}
\partial_{\nu}k^{\mu\nu}_{\xi} = \omega^{\mu}(g;\delta_\xi g,\delta g) + 2\,\xi^{\nu} \delta(E^{\mu\lambda} g_{\lambda\nu}) - \xi^{\mu}\, E^{\nu\lambda}\,\delta g_{\nu\lambda}  ~.
\end{gather}
Using the above results for the terms on the right-hand side, integration by parts leads to $k^{\mu\nu}_{\xi}$ given by
\begin{align}\label{eq:3dkInitial}\nonumber
 k^{\mu\nu}_{\xi} = \frac{\sqrt{-g}}{2\,\kappa^{2}}\, \left(\vphantom{\frac{1}{2}}\right.& \xi^\mu \nabla_\lambda (\delta g)^{\nu\lambda} - \xi^\mu \nabla^\nu (\delta g)^{\lambda}{}_\lambda + \xi_\lambda \nabla^\nu (\delta g)^{\mu\lambda} \\&\left.+ \frac{1}{2} (\delta g)^{\lambda}{}_{\lambda} \nabla^\nu \xi^\mu -\delta g^{\lambda\nu}\nabla_{\lambda}\xi^\mu -(\mu\leftrightarrow\nu) \right) ~.
\end{align}
Note that \eqref{eq:3dkCondition} is satisfied by \eqref{eq:3dkInitial} for all field configurations, independent of whether the equations of motion hold. When the fields are fully on-shell, so that the presymplectic current is conserved, \eqref{eq:3dkCondition} reduces to $\partial_{\nu}k^{\mu\nu}_{\xi} = \omega^{\mu}(g;\delta_{\xi}g,\delta g)$. However, for partially on-shell field configurations only the last term on the right-hand side of \eqref{eq:3dkCondition} vanishes. In that case, some components of $\xi^{\nu} \delta(E^{\mu\lambda} g_{\lambda\nu})$ are non-zero and contribute to $k^{\mu\nu}_{\xi}$.

Evaluating the $k^{ur}_{\xi}$ component of \eqref{eq:3dkInitial} leads to the same problem encountered in section \ref{sec:2d}: it diverges in the limit $r \to \infty$, and the finite part is not integrable. We address this with the prescription outlined in section \ref{sec1:covphasespace} for constructing a finite ($r$-independent) codimension-2 form, and determine the field dependence of the functions appearing in \eqref{AKV} that leads to integrable charges.

\subsection{Charges}

The construction in section \ref{sec1:covphasespacedetail} shifts the components of the codimension-2 form by corner contributions $\Y^{ar}$ from the presymplectic current and $\Z^{ar}$ from the weakly vanishing Noether current, as in \eqref{eq:kShifted}. This defines (up to residual ambiguities) a codimension-2 form such that $\partial_{a}k^{ra}_{\xi}$ is independent of $r$. The charges are obtained from the $ur$ component, which may depend on $r$ through terms whose $u$-derivative cancels against similar terms in $\partial_{\phi}k^{r\phi}_{\xi}$. In other words, any $r$-dependence in $k^{ur}_{\xi}$ is a total $\phi$-derivative and therefore can be discarded in integrals over $\B$ or $\partial\B$. Thus, to determine the charges it is sufficient to calculate only the shift in the $ur$ component of \eqref{eq:3dkInitial}
\begin{gather}
    k^{ru}_{\xi} \to k^{ru}_{\xi} + \Y^{ur}_{\xi} + \Z^{ur}_{\xi} ~.
\end{gather}
However, since we consider field configurations that satisfy the $E^{uu}$, $E^{u\phi}$, and $E^{ur}$ components of the equation of motion, $\Z^{ur}_{\xi} = 0$ and the only shift to the $ur$ component of \eqref{eq:3dkInitial} comes from $\Y^{ur}$. This is obtained from the anti-symmetrized variation of $\Theta^{u}$ written as a total $r$-derivative.

Recall from section \ref{sec1:covphasespace} that the contributions from the $\Theta^{a}$ components to the divergence of the presymplectic current can be folded into the $r$-derivative to define a $\Theta^{r}_\text{ren.}$ that includes corner contributions
\begin{equation}
    \partial_\mu\Theta^\mu =\partial_r\left( \Theta^r + \partial_u \int dr \Theta^u +\partial_\phi \int dr \Theta^\phi \right) := \partial_r\Theta^r_{\text{{ren}}} ~.
\end{equation}
Evaluating the symplectic potential \eqref{eq:3dPresymplecticPotential} on the line element \eqref{eq:3dlineelement} one has 
\begin{align} \label{eq:3dThetauTotalDerivative}
\kappa^2\,\Theta^u&=\partial_r(\delta \mathcal W -\delta\beta\,\,\mathcal W )+2\mathcal W \partial_r\delta\beta \\
\kappa^2\Theta^\phi&= \delta\left( -\frac1{   {\mathcal W} }\partial_\phi e^{2\beta}+\partial_r\W \,\mathcal U\right)+\partial_r\left(-\mathcal W \,\mathcal U\, \delta \beta +\frac12 \mathcal W\, \delta \mathcal U\right)\\ 
&+\frac{2 e^{2\beta}}{ \W } \delta \beta \partial_\phi\beta+ \W \, \partial_r\mathcal U\,\delta\beta+2\W\,\mathcal U\,\partial_r\delta \beta
\end{align}
Since $\partial_{r}\beta = \partial_{r}\delta\beta =0$, $\Theta^{u}$ is automatically a total $r$-derivative in Bondi-Weyl gauge. The $\Theta^{\phi}$ component, on the other hand, must be integrated with respect to $r$. However, as explained above, only the shift coming from the $\partial_{u}\Theta^{u} = \partial_{r} \partial_{u}Y^{ur}$ corner contribution will be relevant for the charges.

Incorporating the corner contribution from $\Theta^{u}$ in the $r$ component of the presymplectic potential, and ignoring both total $\phi$-derivatives (including the contribution from $\Theta^{\phi}$) and $\delta$-exact terms, we have 
\begin{align}\label{eq:3dThetarRenorm}\nonumber
\kappa^2\,\Theta^{r}_{\text{ren}} &=   \kappa^2\,\Theta^{r}+\partial_u\left(\delta \W-\W\,\delta\beta\right)
 \\
 &= - e^{\varphi}(M-\partial_u C-U\,C'+2e^{2\beta-\varphi}\beta'^2)\delta(2\beta-\varphi) \\ \nonumber 
 & \quad +e^\varphi (N-C')\delta U -\partial_u(e^\varphi \delta C) ~. 
\end{align}
The only $r$-dependence takes the form of a $\delta$-exact term, which will not contribute to $\omega^{r}$ and hence has been dropped. Then the $r$ component of the symplectic current is
\begin{align}\nonumber
 \kappa^2\,\omega^{r}_{\text{ren}} 
 & = \kappa^{2}\,\omega^{r} + \partial_{u}\big(\delta_{1}\W\,\delta_{2}\beta - \delta_{2}\W\,\delta_{1}\beta \big) \\
 & = - \delta_1\left(e^{\varphi}(M-\partial_u C-U\,C')+2e^{2\beta-\varphi}\beta'^2\right)\delta_2(2\beta-\varphi)\\&+\delta_1\left(e^\varphi (N-C')\right)\delta_2 U-\partial_u\left(\delta_1e^\varphi \delta_2 C\right)-(1\leftrightarrow2)
\end{align}
The total $u$-derivative on the right-hand side of the first line gives $\Y^{ur}_{\xi}$ when $\delta_{2}$ is a generic field variation, and $\delta_{1} = \delta_{\xi}$ is a field variation with the same form as a diffeomorphism.

Before giving the $ur$ component of the shifted codimension-2 form, let us briefly compare the results above with the two corner terms introduced in \cite{Geiller:2021vpg} to `renormalize' the presymplectic potential. The presymplectic potential in that reference is related to the $\Theta^{r}$ component of \eqref{eq:3dPresymplecticPotential} by
\begin{align}
\Theta^{r\text{\cite{Geiller:2021vpg}}}_\text{ren}&= \Theta^{r} 
- \frac r{\kappa^2}\partial_u(e^\varphi\delta\beta-\delta e^\varphi)+  \frac1{\kappa^2} \partial_u(e^\varphi C\delta \beta) ~.
\end{align}
The first of the two terms canceled the $r$-dependent part of $\Theta^{r}$ that diverges in the $r\to\infty$ limit. The second corner term, which is finite in this limit, led to an integrable codimension-2 form when the functions parameterizing the diffeomorphism \eqref{AKV} have a specific field dependence. This is equivalent to \eqref{eq:3dThetarRenorm}, up to total $\phi$-derivatives and $\delta$-exact terms 
\begin{align}
\Theta^{r}_{\text{ren}}&=\Theta^{r\text{\cite{Geiller:2021vpg}}}_\text{ren} + (\ldots)' + \delta(\ldots) ~.
\end{align}
Thus, the \emph{ad hoc} corner terms introduced in \cite{Geiller:2021vpg} are in fact explained by corner contributions from the component $\Theta^{u}$ of the symplectic potential. Note the second corner term introduced in \cite{Geiller:2021vpg}, which is necessary for integrability of the charges, appears as the $r$-independent part of $Y^{ur}$ when we use integration-by-parts to express $\Theta^{u} = \partial_{r} Y^{ur}$ as a total $r$-derivative.

Incorporating the $\Y^{ur}_{\xi}$ shift to the $ur$ component of \eqref{eq:3dkInitial}, and dropping total $\phi$-derivatives, $k^{ur}_{\xi}$ is given by
\begin{align}
    k^{ur}_{\xi} = & \,\, f\,\Big[ e^{2\beta}\,\delta\Big(-2e^{-2\beta}\,\big(e^{2\beta-\varphi}\beta' \big)'\Big) + e^{2\beta-\varphi}\delta\big(-M\,e^{2\varphi-2\beta}\big) \\ \nonumber
    & \qquad + e^{2\beta}\,\delta\Big(U\,e^{\varphi-2\beta} C' \Big) + e^{2\beta}\,\delta\Big(e^{\varphi-2\beta}\partial_{u}C\Big)\Big] \\ \nonumber
    & \,\, + \Big(-k + e^{2\beta-2\varphi}\,\big(f'' + f'\,(4\beta-\varphi)'\big)\Big)\delta(e^{\varphi}) \\ \nonumber
    & \,\, + \Big(h + U\,f' - g'-g\,\varphi'-f\,\partial_{u}\varphi\Big)\,e^{\varphi}\,\delta C \\ \nonumber
    & \,\, + \big(f\,U - g\big)\,\delta\big(e^{\varphi}\,N\big) + f\,\delta C\,\Big(\partial_{u} e^{\varphi} + \big(U\,e^{\varphi}\big)'\Big) ~.
\end{align}
This becomes integrable when the field dependence of the functions in the asymptotic Killing vectors is
\begin{align}
f&=\bar{f}e^{\varphi-2\beta},\qquad h=-\big(\bar{h}+(\bar{g}e^\varphi)'\big)e^{-\varphi}+\bar{g}'+\bar{g}\varphi',\\
g&=\bar{g}+\bar{f}e^{\varphi-2\beta}U,\qquad
k=\bar{k}-\bar{g}C'-\bar{f}e^{\varphi-2\beta}(UC'+\partial_u C)\,.
\end{align}
The functions $\bar{f}$, $\bar{g}$, $\bar{h}$, and $\bar{k}$ are taken to be independent of the fields, but are otherwise arbitrary functions of $(u,\phi)$. With this assignment, the charges are then precisely the ones presented in \cite{Geiller:2021vpg}:
\be\label{final charge}
\delta Q_\xi=\int_{0}^{2\pi} \!\!\!\!\de \phi\,\Big(\bar{f}\,\delta\bar{M}+\bar{g}\,\delta\bar{N}+\bar{h}\,\delta C+\bar{k}\,\delta e^\varphi \Big) \, ,
\ee
with
\begin{gather}\label{eq:barMcharge}
\bar{M}=4(\beta')^2-2\beta'\varphi'+\f{1}{2}(\varphi')^2+(2\beta-\varphi)''+e^{2(\varphi-\beta)}\big(M-\partial_u C {- U\,C'}\big) \\ \label{eq:barNcharge}
\bar{N}=e^\varphi\big(N-C'\big).
\end{gather}
When $\delta \bar f=\delta\bar g=\delta\bar h=\delta\bar h=0$, the algebra of charges is 
\begin{equation}
[Q_{\bar\xi_1},Q_{\bar\xi_2}]=Q_{[\bar\xi_1,\bar\xi_2]_\star}+ \int_{0}^{2\pi}\!\!\!\!\de \phi \Big(\bar{f}_1\bar{g}_2'''-\bar{f}_2\bar{g}_1''' + \bar{h}_1\bar{k}_2-\bar{h}_2\bar{k}_1 \Big),
\end{equation}
where $[\bar\xi_1,\bar\xi_2]_\star=\xi(\bar{f}_{12},\bar{g}_{12},\bar{h}_{12},\bar{k}_{12})$ is the vector 
with
\bsub\label{commutation relations new slicing}
\be
\bar{f}_{12}&=\bar{f}_1\bar{g}_2'+\bar{g}_1\bar{f}_2'-(1\leftrightarrow2),\qquad \bar{h}_{12}=0\\
\bar{g}_{12}&=\bar{g}_1\bar{g}'_2+\f{1}{\ell^2}\bar{f}_1\bar{f}_2'-(1\leftrightarrow2)
\,,\qquad 
\bar{k}_{12}=0\,.
\ee
\esub
Explicitly this is an algebroid whose base space is parametrized by $u$ and at each $u$ we have the conformal algebra in two dimensions in direct sum with the Heisenberg algebra. In the flat limit the conformal algebra becomes the BMS algebra in three dimensions.

\subsection{First variation of the action and holographic renormalization}

The goal of this section is to use what we learned about the structure of the presymplectic form in the previous sections to construct an action with a well-defined variational principle, for fields that satisfy Dirichlet boundary conditions at $r \to \infty$. In Bondi-Weyl gauge \eqref{eq:3dlineelement} with the partially on-shell fields \eqref{W solution}-\eqref{V solution}, we take Dirichlet boundary conditions to fix the functions $\beta$, $\varphi$, $U$, and $C$ characterizing the large-$r$ behavior of the metric. The first variation of the action should vanish for variations of the fields that preserve this boundary data.  The functions $M$ and $N$, which are sub-leading in the large $r$ expansion of the metric, are not fixed.

Since $E^{\mu\nu}\,\delta g_{\mu\nu} = 0$ for these field configurations, the variation of the bulk Lagrangian is just the total derivative $\delta L_{M} = \partial_{\mu}\Theta^{\mu}$ with support on $\partial M$. Our focus is the contribution on the constant-$r$ surface $\B$ that becomes spatial infinity in the limit $r \to \infty$. The $\Theta^{r}$ component of the presymplectic potential contains factors that diverge as $r \to \infty$. As a result, it is non-zero for some field variations that preserve the boundary conditions. This is most easily seen by considering field variations that change the boundary data; i.e. field variations $\delta \beta$, $\delta \varphi$, $\delta U$, and $\delta C$. Such variations produce terms at $\O(r^2)$ and $\O(r)$ in $\Theta^{r}$. The $\O(r^2)$ term is $\delta$-exact, so it could in principle be canceled by a suitable boundary term added to the action. But the $\O(r)$ term is \emph{not} $\delta$-exact. So the terms in $\Theta^{r}$ that interfere with a well-defined variational principle cannot be entirely removed by boundary terms in the action.

However, as we saw in the previous section, contributions from the $\Theta^{a}$ components must be accounted for. They can be expressed as total $r$-derivatives, $\Theta^{a} = \partial_{r} Y^{ar}$, which in turn give corner contributions on $\B$. If we consider the shifted quantity $\Theta^{r} + \partial_{a}Y^{ar}$, which represents all contributions on $\B$ (including $\partial \B$) from the variation of the bulk Lagrangian, then field variations which shift the boundary data produce $\delta$-exact terms at both $\O(r^2)$ and $\O(r)$. 
\begin{gather}
    \Theta^{r} + \partial_{a}Y^{ar} = -\frac{1}{\ell^{2}}\,\delta\Big(e^{2\beta-\varphi}\,\W^{2}\Big) + \big(\text{terms independent of $r$}) ~.
\end{gather}
The obstruction to a well-defined variational principle is now a $\delta$-exact term which can be addressed by the variation of boundary terms; this is the correct starting point for constructing an action with a well-defined variational principle. All that remains is to identify suitable boundary terms such that $\delta \Gamma = 0$ for field variations that preserve the Dirichlet boundary data.

Before presenting the action and verifying its properties, we comment on two ambiguities present in these constructions. First, as we saw with the two-dimensional dilaton gravity models of section \ref{sec:2d}, it is natural to consider terms in the action with support on the null components of the boundary $\N_{\pm}$. We always work with an affine parameterization of the null generators on $\N_{\pm}$, so that these terms and their first variation are zero except for contributions that arise at the corners $\B \cap \N_{\pm}$ \cite{Hayward:1993my, Lehner:2016vdi}. Second, given an action with a well-defined variational principle, one can define another such action by adding an intrinsic functional of the boundary data on $\B$ that remains finite as $r \to \infty$ (assuming such a functional of the fields exists). Both actions have the same bulk equations of motion. And since the new term vanishes for field variations which preserve the boundary data there is no impact on the variational principle. Here, we fix (or at least reduce) this ambiguity by demanding that the action have a finite flat-space limit $\ell \to \infty$ for partially on-shell fields in Bondi-Weyl gauge, without any additional rescalings or redefinitions of the fields.

Given the considerations outlined above, we arrive at an action that admits a well-defined variational principle with Dirichlet boundary conditions, is finite in the $r \to \infty$ limit that removes $\B$ to spatial infinity, and has a sensible flat space limit $\ell \to \infty$. As expected, it is similar in form to the usual action for AAdS$_3$ gravity, with additional terms needed in the presence of the corners $\partial \B$ and to accommodate the $\ell \to \infty$ limit:
\begin{align}\label{eq:3dActionWithCornersStandardTerms}
    \Gamma = & \,\,  \frac1{2\kappa^{2}} \int_{M} \nts d^{3}x \sqrt{-g}\,\left(R+ \frac{2}{\ell^2} \right)  + \frac{1}{\kappa^2}\,\int_{\B}d^{2}x \sqrt{-\gamma} \left( K-\frac1{\ell} \right) \\ \label{eq:3dActionWithCornersNewTerms}
    & + \frac{1}{\kappa^2}\,\int_{\B}d^{2}x \sqrt{-\gamma} \left(\f{\ell}2\,\,K_{c}^2 \right) + \frac{1}{\kappa^{2}} \,\int_{\partial \B}\!\! d^{1}x  \, \sqrt{\sigma} \,\big(-\ell\, K_{c} -1 \big) ~.
\end{align} 
Here $K = \nabla_{\mu} n^{\mu}$ is the trace of the extrinsic curvature of the constant-$r$ surface $\B$ embedded in $M$ with outward-pointing, spacelike unit normal $n^{\mu}$. The coordinates on $\B$ are $x^{a}=(u,\phi)$, and $\gamma_{ab}$ is the induced metric. In addition, $\B$ is foliated into curves of constant $u$, with forward-pointing timelike unit normal $\rho^{a}$. The extrinsic curvature of the leaves in this foliation is $K_{c} = D_{a}\rho^{a}$, where $D_{a}$ is the covariant derivative on $\B$ compatible with the metric $\gamma_{ab}$. The one-dimensional metric along each curve in the foliation is $\sigma\,\de \phi^{2}$.

The terms \eqref{eq:3dActionWithCornersStandardTerms} in the first line of the action are the familiar bulk and boundary terms for Einstein-Hilbert gravity in three dimensions, with asymptotically anti-de Sitter boundary conditions. They include the Gibbons-Hawking-York (GHY) term and a boundary counterterm \cite{Henningson:1998gx, Balasubramanian:1999re, Emparan:1999pm}. The second line \eqref{eq:3dActionWithCornersNewTerms} includes new terms that are needed to implement the variational principle when $\B$ has corners, and to ensure a finite $\ell \to \infty$ limit.

\paragraph{First variation of the action}
The variational properties of the terms in \eqref{eq:3dActionWithCornersStandardTerms} are well established when $\partial M$ does not have corners. To verify our claims about the full action it is sufficient to show that the contributions at the corners $\partial \B$ do not interfere with the variational principle, and that the $\ell \to \infty$ limit is well defined.

First, consider the terms in $\delta \Gamma$ with support on $\B$ coming from the first line \eqref{eq:3dActionWithCornersStandardTerms}.
\begin{gather}\label{eq:3dFirstVariationStandardTerms}
    \delta \Gamma = \ldots + \int_{\B}\!\! \de^{2}x\,\left[\Theta^{r} + \partial_{a}Y^{ar} + \frac{1}{\kappa^{2}}\,\delta\,\left(\sqrt{-\gamma}\,K - \sqrt{-\gamma}\,\frac{1}{\ell}\right)\right]
\end{gather}
The variation of the GHY term and boundary counterterm gives
\begin{align}
 \delta \Big(L_\text{GHY} + L_\text{CT}\Big) = &\,\, -\Theta^{r} + \frac{1}{2\,\kappa^{2}}\sqrt{-\gamma}\,\left(\gamma^{ab}\Big(K - \frac{1}{\ell}\Big) - K^{ab}\right) \delta\gamma_{ab} \\ \nonumber 
   & \quad + \frac{1}{\kappa^{2}}\,\sqrt{-\gamma}\,D_{a} c^{a} ~.
\end{align}
This includes a commonly neglected total derivative term arising from the variation of the extrinsic curvature \cite{FormulasWebsite}, which cannot be ignored in the presence of corners $\partial \B$. As above, $D$ is the two-dimensional covariant derivative on $\B$ compatible with $\gamma$, and the vector $c^{a}$ is given by
\begin{gather}\label{eq:3dcaDefinition}
    c^{a} = - \frac{1}{2}\,\gamma^{ab}\,\delta g_{b\lambda}\,n^{\lambda} ~.
\end{gather}
Together, the terms in \eqref{eq:3dFirstVariationStandardTerms} reduce to a familiar contribution on $\B$ and a corner term with support on $\partial \B$
\begin{gather}\label{eq:PartialdeltaGamma}
    \delta \Gamma = \ldots + \int_{\B}\!\! \de^{2}x\,\left[\frac{\sqrt{-\gamma}}{2\,\kappa^{2}}\,\left(\gamma^{ab}\Big(K - \frac{1}{\ell}\Big) - K^{ab}\right)\delta \gamma_{ab} + \partial_{a}\Big(Y^{ar} + \frac{1}{\kappa^{2}}\,\sqrt{-\gamma}\,c^{a} \Big)\right] ~.
\end{gather}
The first set of terms, involving the extrinsic curvature and the variation of the induced metric on $\B$, vanishes for field variations that preserve Dirichlet boundary conditions at $r \to \infty$ \cite{Papadimitriou:2005ii}. But this is not the case for the corner term. From \eqref{eq:3dThetauTotalDerivative} we have $Y^{ur} = \delta\W - \W\,\delta\beta$, and the definition \eqref{eq:3dcaDefinition} gives $\sqrt{-\gamma}\,c^{u} = \W\,\delta\beta$. Thus, the variation of the first line of the action \eqref{eq:3dActionWithCornersStandardTerms} includes a corner term that interferes with the definition of the variational principle. For field variations that change the boundary data it takes the form
\begin{gather}\label{eq:3dProblematicCornerTerm}
    \frac{1}{\kappa^{2}} \int_{\partial\B}\!\! d\phi\,\delta \W  ~.
\end{gather}
This diverges as $r \to \infty$, since $\W = e^{\varphi}\,\big(r - C\big)$. Therefore, there will be field variations that preserve the boundary data for which the corner terms in \eqref{eq:PartialdeltaGamma} are non-zero. This is at odds with our requirements for the variational principle.

The corner term described above, which is $\delta$-exact, is addressed by adding the very last term in \eqref{eq:3dActionWithCornersNewTerms} to the action. It takes the form
\begin{gather}
    - \frac{1}{\kappa^{2}} \int_{\partial\B}\!\! d^{1}x\,\sqrt{\sigma} ~.
\end{gather}
A leaf of the foliation of $\B$ by curves of constant $u$ has induced metric $\sigma\,\de\phi^{2}$, so $\sqrt{\sigma} = \W$ for the fields \eqref{W solution}. Thus, the variation of this term in the action immediately cancels the problematic term \eqref{eq:3dProblematicCornerTerm}.

All that remains in our proposed action are the terms in \eqref{eq:3dActionWithCornersNewTerms} involving the extrinsic curvature $K_c$ of the constant-$u$ curves foliating $\B$. It is straightforward to see that these terms are a finite (as $r \to \infty$) functional of the Dirichlet boundary data on $\B$. The future-pointing timelike unit vector normal to curves of constant $u$ on $\B$ is
\begin{gather}
    \rho^{a} = \frac{1}{e^{\beta}\,\sqrt{-\V}}\,\Big(\delta^{a}{}_{u} + \U\,\delta^{a}{}_{\phi} \Big) ~,
\end{gather}
so the extrinsic curvature of these curves is
\begin{gather}
    K_{c} = D_{a}\rho^{a} = \frac{1}{\sqrt{-\gamma}}\,\Big(\partial_{u}\W + \partial_{\phi}\big(\W\,\U\big) \Big) ~.
\end{gather}
From the large-$r$ behavior of the fields \eqref{W solution}-\eqref{V solution} it is apparent that the quantities $\sqrt{-\gamma}\,K_{c}^{2}$ and $\sqrt{\sigma}\,K_c$ in \eqref{eq:3dActionWithCornersNewTerms} are finite functions of the Dirichlet boundary data as $r \to \infty$, and hence do not interfere with the variational principle. An alternate presentation of these terms is given in Appendix \ref{app:geometry}.

Including the contributions from all terms in \eqref{eq:3dActionWithCornersStandardTerms}-\eqref{eq:3dActionWithCornersNewTerms}, and evaluating on fields that satisfy $E^{\mu\nu}\,\delta g_{\mu\nu} = 0$, we have $\delta \Gamma = 0$ for all field variations that preserve the Dirichlet boundary data at $r \to \infty$. For field variations that change this boundary data, the non-zero terms in $\delta \Gamma$ are
\begin{align} \label{eq:3dfirstvariation}
    \delta \Gamma  = \frac1{\kappa^2}\int_{\B} \nts du d\phi  \bigg[&\partial_u(-e^{\varphi}\delta C)+e^\varphi \delta \left(2\beta-\varphi\right) \left(-M+\partial_u C+U\,C'\right)\\ \nonumber & +e^\varphi \delta U \left( N-C' \right)  -\delta\beta \left(4e^{2\beta-\varphi}\beta' \right) '\bigg] ~.
\end{align} 
Total derivatives with respect to $\phi$ have been dropped to obtain this form of $\delta\Gamma$. It is interesting to note that this can also be written as
\begin{align} \label{eq:3dfirstvariation2}
    \delta \Gamma = \frac1{\kappa^2}\int_{\B} \nts du d\phi  \bigg[&\partial_u(-e^{\varphi}\delta C)-\bar M\,\delta(e^{2\beta-\varphi} ) +  \bar N\,\delta U  \\ \nonumber &+ \delta\left(-2\left(e^{2\beta-\frac12\varphi}\right)' \left(e^{-\frac12\varphi}\right)' \right) \bigg]~,
\end{align}
where $\bar{M}$ and $\bar{N}$ are the charges defined in \eqref{eq:barMcharge}-\eqref{eq:barNcharge}.

\paragraph{On-shell action}

The value of the action for a specific field configuration will depend on details in the interior of $M$, like the presence of a horizon. Evaluating \eqref{eq:3dActionWithCornersStandardTerms}-\eqref{eq:3dActionWithCornersNewTerms} for the fields \eqref{W solution}-\eqref{V solution} on a semi-infinite interval $r_{+} \leq r < \infty$, gives
\begin{align}
\Gamma = &\frac1{\kappa^2} \int_{\B}\de u \de \phi\,\left( 
    -e^\jf \,\big(M - \partial_u C - U\,C' \big) + \frac{1}{\ell^{2}}\,e^{2\beta + \varphi}\big(r_{+} - C\big)^{2}\right) ~.
\end{align} 
Total $\phi$-derivatives have been discarded, and the last term captures contributions from the lower limit of the integral over $r$. This result makes two things apparent. First, the boundary terms needed for a well-defined variational principle also render the action finite as the regulator is removed via the $r \to \infty$ limit. And second, the action remains finite in the flat space $\ell \to \infty$ limit.

\section{Conclusion}

In this paper we have outlined a procedure for obtaining finite charges at an asymptotic boundary, and applied it to two lower-dimensional examples where the resulting charges are integrable. The existence of the limiting procedure that defines the charges is ensured by appropriate corner terms in the presymplectic current, obtained directly from the variation of the bulk Lagrangian. This method of identifying the corner terms gives a systematic accounting of the \emph{ad hoc} shifts and subtractions used in some previous calculations \cite{Ruzziconi:2020wrb,Geiller:2021vpg,Campoleoni:2022wmf,Adami:2023fbm}. In these lower-dimensional examples, the charges are not just finite at the asymptotic boundary, but fully independent of the coordinate $r$ used to implement the limiting procedure. One major advantage of this approach is that it reduces ambiguities in Iyer-Wald covariant phase space methods. A (finite) ambiguity remains, which is to be expected in any formulation that accommodates an open system where the boundary dynamics is not specified. We also show that the corner terms in the presymplectic potential provide important information needed to construct a well-defined variational principle on spacetimes where components of the boundary intersect at corners.

A notable feature of our analysis is that the fields are not fully on shell. Some equations of motion are enforced, while others -- constraints conjugate to gauge-fixed quantities -- are relaxed. Nevertheless, even with the fields only ``partially on-shell,'' diffeomorphism invariance implies the existence of a codimension-2 form which receives contributions from both the presymplectic current and the weakly vanishing Noether current. When the fields are fully on-shell it reduces to the usual result, but with our weaker conditions it captures charges which are not otherwise accessible. This formalizes some of the assumptions in the analysis of \cite{Grumiller:2017qao}, where on-shell conservation of the Casimir must be relaxed before the off-shell conformal symmetry is recovered in the dilaton gravity dual of the SYK model.

The examples in sections \ref{sec:2d} and \ref{sec:3d} are lower-dimensional theories without local degrees of freedom in the bulk. A natural next step is to apply the procedure to higher dimensional theories with dynamical degrees of freedom that might radiate across the asymptotic boundary. The proof outlined in section \ref{sec1:covphasespace} will still apply if the fields fall off sufficiently fast as $r\to \infty$. The question now is whether those fall-off conditions are physically acceptable. One starting point is four-dimensional gravity in the partial Bondi gauge studied in \cite{Geiller:2022vto}, which is closely related to the three-dimensional Bondi gauge used in section \ref{sec:3d}. The expectation is that the components $\Theta^{a}$ at large $r$ will then consist of a leading part fixed by the boundary conditions and kinematics, along with a sub-leading part where the dynamical degrees of freedom appear. The Iyer-Wald ambiguity would presumably be used to shift the former into corner contributions appearing in $\Theta_\text{ren}^{r}$. This may be sufficient to define charges which satisfy an appropriate flux-balance law \cite{Wald:1999wa,Barnich:2011mi,Compere:2018ylh,Compere:2019gft}.
However, it is also possible that the remaining, dynamical part of $\omega_\text{ren}^{u}$ does not fall off fast enough at large $r$ for the arguments of section \ref{sec1:covphasespace} to apply. In that case, one would have to either relax the boundary conditions, or else modify our prescription. Since a number of possible complications are expected to appear in four dimensions, it might instead be helpful to start by coupling additional fields to the simple models treated in this paper. One could then focus on just the issue of extracting corner terms when dynamical degrees of freedom appear in the relevant components of the presymplectic potential.

Even without adding new fields, there are still important results to recover in the lower-dimensional models we consider. For three dimensional gravity, one expects as many as six charges \cite{Grumiller:2016pqb,Grumiller:2017sjh}. The Bondi-Weyl gauge employed in section \ref{sec:3d} leads to four charges. 
Relaxing the $g_{r\phi} = 0$ condition would bridge the Bondi-Weyl \cite{Geiller:2021vpg} and covariant Bondi \cite{Campoleoni:2022wmf} gauges, enlarging the space of field configurations and possibly increasing the number of charges. As shown in sections \ref{sec:2d} and \ref{sec:3d}, contributions from the variation of the weakly vanishing Noether current in \eqref{eq:SecondCodimension2Definition} are essential for obtaining the charges when the fields are only partially on-shell. Removing the gauge condition $g_{r\phi} = 0$ would allow variations $\delta g_{r\phi}$, and therefore the $E^{r\phi} = 0$ component of the equation of motion would presumably need to be enforced to completely fix the $r$-dependence of the fields. This will change how the weakly vanishing Noether current contributes to the charges.

There are other extensions and refinements of this work that we would like to explore. The focus here has been on asymptotic symmetries generated by diffeomorphisms. It would be useful to generalize our analysis to incorporate other gauge symmetries \cite{Barnich:2005kq, Compere:2007vx, Compere:2007az, Barnich:2015jua, Detournay:2018cbf, Spindel:2018cgm}. Even the addition of a simple $U(1)$ gauge field would be interesting, since unlike a scalar it would contribute to the weakly vanishing Noether current. Another avenue of investigation involves the choice of spacelike coordinate $r$ that is singled out when specifying the boundary conditions and constructing the charges. For some theories, such as asymptotically AdS gravity, the boundary conditions can be formulated in a completely covariant manner \cite{Ashtekar_Magnon_1984, Ashtekar_Das_2000}. It seems reasonable that the procedure for extracting the corner terms in the presymplectic potential could also be specified in a covariant manner for such theories.

\section*{Acknowledgments}
The authors thank Adrien Fiorucci for helpful discussions and comments on a draft of this paper. We also thank Daniel Grumiller and Florian Ecker for stimulating conversations, and Technical University of Vienna for hospitality during the early stages of this project. Some of the calculations presented in this paper were performed with the {\tt xAct} package for {\sc Mathematica} \cite{MartinGarcia}. Research at Perimeter Institute is supported in part by the Government of Canada through the Department of Innovation, Science and Economic Development Canada and by the Province of Ontario through the Ministry of Colleges and Universities.

\appendix

\section{``Einstein-Hilbert'' contribution to $k_{\xi}^{\mu\nu}$ in 2d dilaton gravity}
\label{app:kEH}
The last line in \eqref{eq:kInitial} has been simplified. It is proportional to the product of the dilaton $X$ and the codimension-2 form $k^{\mu\nu}_{\xi, EH}$ for Einstein-Hilbert gravity, which is given by 
\begin{align}\label{eq:kEH1}
    \frac{2\,\kappa^{2}}{\sqrt{-g}}\,k_{\xi,EH}^{\mu\nu} = & \,\,  \xi^{\nu}\,\nabla^{\mu}\delta g - \xi^{\mu}\,\nabla^{\nu}\delta g + \xi^{\mu}\,\nabla_{\lambda} (\delta g)^{\lambda\nu} - \xi^{\nu}\,\nabla_{\lambda} (\delta g)^{\mu\lambda} \\ \nonumber
    & + \xi^{\lambda}\,\big(\nabla^{\nu}(\delta g)^{\mu}{}_{\lambda} - \nabla^{\mu}(\delta g)_{\lambda}{}^{\nu}\big) + \frac{1}{2}\,\delta g\,\big(\nabla^{\nu}\xi^{\mu} - \nabla^{\mu}\xi^{\nu}\big) \\ \nonumber
    \vphantom{\frac{2\,\kappa^{2}}{\sqrt{-g}}} &  + (\delta g)^{\mu\lambda}\,\nabla_{\lambda}\xi^{\nu} - (\delta g)^{\lambda\nu}\,\nabla_{\lambda}\xi^{\mu} ~.
\end{align}
The terms in the last line of \eqref{eq:kInitial} (with a factor of $\sqrt{-g}/\kappa^{2}$) are equal to $X\,k^{\mu\nu}_{\xi, EH}$ in two dimensions. This is not immediately apparent when the two expressions are compared in covariant form, and the condition \eqref{eq:kCondition} satisfied by $k^{\mu\nu}_{\xi}$ is most easily verified by replacing the last line of \eqref{eq:kInitial} with $X\,k^{\mu\nu}_{\xi, EH}$. We include it here for completeness. In any case, this part of the codimension-2 form does not contribute to our charge calculation. 
Evaluating \eqref{eq:kEH1} in linear dilaton Bondi gauge gives
\begin{gather}
    k_{\xi,EH}^{ur} = \frac{1}{2\kappa^{2}}\,\delta\big(\sqrt{-g}\,g^{rr}\big)\,\partial_{r}\xi^{u} ~.
\end{gather}
This vanishes for diffeomorphisms that preserve the gauge condition $g_{rr}=0$, which requires $\partial_{r}\xi^{u} = 0$.

\section{Counterterms for dilaton gravity}\label{app:2dhamiltonjacobi}

In this appendix we generalize the results of \cite{Grumiller:2007ju} to dilaton gravity models where spatial infinity is the $r_c \to \infty$ limit of a regulating boundary $\B$ that is not an isosurface of the dilaton. Under certain assumptions the resulting terms can be organized in a derivative expansion on $\B$, with the zeroth-order term reproducing the boundary counterterm given in \cite{Grumiller:2007ju}. Depending on the potentials $U$ and $V$, additional terms in this expansion may be necessary for a well-defined variational problem and a finite action. 

The approach of \cite{Grumiller:2007ju} identified a boundary counterterm for the dilaton gravity action by solving the radial Hamiltonian constraint. That analysis assumed the fields were fully on-shell, and used coordinates $\tilde{r}$ and $t$ adapted to the Killing vector $\xi_{X}$. Curves of constant $\tilde{r}$, including spatial infinity at $\tilde{r} \to \infty$, are then isocurves of the dilaton $X(\tilde{r})$. Here we relax this condition and repeat the analysis for a more general foliation. The starting point is a ``radial ADM'' decomposition of the 2D metric\,\footnote{The linear Bondi gauge is obtained by taking $\gamma_{uu}=2 B, N^u=-e^A/(2B), N=e^A/\sqrt{-2B}$}
\begin{gather}
    ds^{2} = N^{2}\,dr^{2} + \gamma_{uu}\,\big(du + N^{u}\,dr\big)^{2} ~.
\end{gather}
The bulk Lagrangian \eqref{eq:BulkLagrangianDG} can be written as
\begin{align}
    \kappa^{2}\,L_{M} = &\,\, \sqrt{-\gamma}\,\frac{1}{2N}\,\gamma^{uu}\big(\partial_{r}\gamma_{uu} - 2\gamma_{uu} D_{u}N^{u}\big)\big(\partial_{r}X - N^{u} D_{u} X\big) \\ \nonumber
    & \quad -\sqrt{-\gamma}\,\frac{1}{2N}\,U\,\big(\partial_{r}X - N^{u} D_{u} X\big)^2 - \sqrt{-\gamma}\,N\,\Big(\frac{1}{2}\,U\,(DX)^2 + D^{2}X + V \Big) \\ \nonumber& \quad + \partial_{u}\Big(-\sqrt{-g}\,n^{u}\,X\,K + \sqrt{-\gamma}\,\gamma^{uu}\,\big(N\,\partial_{u}X - X\,\partial_{u}N\big)\Big) \\ \nonumber & \quad + \partial_{r}\Big(-\sqrt{-g}\,n^{r}\,X\,K\Big) ~,
\end{align}
where $K$ is the extrinsic curvature of a constant $r$ curve, $\gamma_{uu}$ is the induced metric on that curve, and $D_{u}$ is the (1D) covariant derivative compatible with that metric. The last term cancels the GHY term on $\B$ so we drop it henceforth.~\footnote{That term potentially gives another contribution if the boundary has a component at some interior value of $r$, like a horizon. But that is not relevant here, as we are only interested in terms on $\B$.} The momenta conjugate to $\partial_r \gamma_{uu}$ and $\partial_{r}X$ are
\begin{align}
    \pi^{uu} = &\,\, \frac{\partial L_{M}}{\partial(\partial_{r}\gamma_{uu})} = \frac{\sqrt{-\gamma}}{2\,\kappa^{2} N}\,\gamma^{uu}\big(\partial_{r}X - N^{u} D_{u} X\big) \\
    \pi_{X} = &\,\, \frac{\partial L_{M}}{\partial(\partial_{r}X)} = \frac{\sqrt{-\gamma}}{2\,\kappa^{2} N}\,\gamma^{uu}\big(\partial_{r}\gamma_{uu} - 2\,\gamma_{uu} D_{u} N^{u}\big) - 2\,U\,\pi^{u}{}_{u} ~.
\end{align}
Now we construct the Hamiltonian $H_{M} = \pi^{uu}\,\partial_{r}\gamma_{uu} + \pi_{X}\,\partial_{r}X - L_{M}$. Using the above expressions, this is
\begin{gather}
    H_{M} = N\,\H + N^{u}\,\H_{u} + \partial_{u} T^{u} ~,
\end{gather}
with
\begin{align}
    \H = &\,\, \frac{2\kappa^{2}}{\sqrt{-\gamma}}\,\Big(\pi_{X}\pi^{u}{}_{u} + U\,(\pi^{u}{}_{u})^{2}\Big) + \frac{\sqrt{-\gamma}}{\kappa^{2}}\,\Big(\frac{1}{2}\,U\,(D X)^2 + D^{2}X + V \Big) \\
    \H_{u} = &\,\, \pi_{X}\,\partial_{u}X - 2\,\partial_{u}\pi^{u}{}_{u} + \pi^{uu}\partial_{u}\gamma_{uu} \\ 
    T^{u} = &\,\, N^{u}\Big(2(1- X\,U) \pi^{u}{}_{u} - X\,\pi_{X}\Big) - \frac{\sqrt{-\gamma}}{\kappa^{2}}\gamma^{uu}\Big( N\,\partial_{u}X - X\,\partial_{u} N\Big) ~.
\end{align}
Now consider a boundary term $L_{\text{{CT}}}(\gamma_{uu},X)$ that is a function of the induced metric and dilaton on $\B$. Defining 
\begin{gather}
    p_{X} = \frac{\delta L_{\text{{CT}}}}{\delta X} \qquad p^{uu} = \frac{\delta L_{\text{{CT}}}}{\delta \gamma_{uu}} ~,
\end{gather}
the goal is to determine $L_\text{CT}$ by solving $\H(p_{X},p^{uu}) = 0$ order-by-order in a derivative expansion. The Hamilton-Jacobi equation for $L_{\text{{CT}}}$ is
\begin{gather}
    \H(p_{X},p^{uu}) = \frac{2\kappa^{2}}{\sqrt{-\gamma}}\,\Big(p_{X}p^{u}{}_{u} + U\,(p^{u}{}_{u})^{2}\Big) + \frac{\sqrt{-\gamma}}{\kappa^{2}}\,\Big(\frac{1}{2}\,U\,(D X)^2 + D^{2}X + V \Big) = 0~.
\end{gather}
Since $L_\text{CT}$ is a scalar on $B$, the first few terms in the derivative expansion should take the form
\begin{gather}\label{eq:LctAnsatz}
    L_{\text{{CT}}} = {\frac1{\kappa^2}}\sqrt{-\gamma}\,\left( F_{0}(X) + F_{2}(X)\,\gamma^{uu}\,(\partial_u X)^{2} + \ldots \right) ~.
\end{gather}
Taking the functional derivatives of $L_{\text{{CT}}}$ and plugging the resulting $p_{X}$ and $p^{u}{}_{u}$ into $\H$, the zero-derivative term in the equation is solved by the counterterm found in \cite{Grumiller:2007ju}:
\begin{gather}
    F_{0} = -\sqrt{e^{-Q(X)}\,w(X)} ~,
\end{gather}
where $Q(X)$ and $w(X)$ were defined in \eqref{eq:Qw}. Using this result for $F_{0}$ in \eqref{eq:LctAnsatz}, the two-derivative term in $\H = 0$ becomes
\begin{align}
    0 = &\,\,\Big(1 + 2\,F_{2}(X)\,\sqrt{e^{-Q}w}\Big)\,D^{2}X \\ \nonumber 
    & \quad - \frac{1}{2}(D X)^{2}e^{-Q}\,\Big(-2\partial_{X}\big(F_{2}(X)e^{Q}\sqrt{e^{-Q}w}\big) - \partial_{X} e^{Q} \Big) ~,
\end{align}
which is solved by
\begin{gather}
    F_{2}(X) = - \frac{1}{2\,\sqrt{e^{-Q(X)} w(X)}} ~.
\end{gather}
Subsequent terms in the derivative expansion of $\H$ involve contributions from terms in $L_{\text{{CT}}}$ with four or more derivatives. So, to second order in the derivative expansion, $L_\text{CT}$ is given by 
\begin{gather}\label{eq:GeneralLct}
    L_{\text{{CT}}} = -{\frac1{\kappa^2}}\sqrt{-\gamma}\left(\sqrt{e^{-Q}w} + \frac{1}{2\,\sqrt{e^{-Q}w}}\,\gamma^{uu} (\partial_{u}X)^2 + \ldots \right) ~,
\end{gather}
where $\ldots$ denotes terms with four or more derivatives. This generalizes the result of \cite{Grumiller:2007ju} and reproduces, for instance, the boundary counterterm used for the Jackiw-Teitelboim model \cite{Jackiw:1984, Teitelboim:1984} in \cite{Grumiller:2017qao}.

Under what conditions can terms with four or more derivatives in \eqref{eq:GeneralLct} be neglected? In linear dilaton Bondi gauge, consider models for which $w(X)$ dominates the $\partial_{u}X$ term in the metric function $B$ as $r \to \infty$. This condition, $(\partial_{u}X)/w \to 0$ as $r \to \infty$, ensures that the two-derivative term in \eqref{eq:GeneralLct} is sub-leading compared to the zero-derivative term. Then the $r \to \infty$ expansion of the counterterm is
\begin{gather}\label{eq:LCTasymptotic}
 {\kappa^2}   L_\text{CT} \sim -w(X)e^{a+q} - \partial_{u}X + e^{a+q}\M + \frac{(\partial_{u}X)^{2}}{e^{a+q}w(X)} + \ldots ~.
\end{gather}
The last term in this expression vanishes as $r \to \infty$ if the potentials $U$ and $V$ give $w(X)$ that grows faster than $X^{2}$. In that case, only the zero-derivative term in \eqref{eq:GeneralLct} contributes to the action and its first variation in this limit. However, for the Jackiw-Teitelboim model \cite{Jackiw:1984, Teitelboim:1984} and related theories \cite{Witten:2020ert} with $w(X) \sim X^{2}$ as $r \to \infty$, the last term in \eqref{eq:LCTasymptotic} is finite and non-zero as $r \to \infty$. In that case the two-derivative term in \eqref{eq:GeneralLct} is not needed to address $r \to \infty$ divergences in the analysis of the variational principle. But for Jackiw-Teitelboim gravity this term's finite contribution to the action \textit{is} necessary for reproducing the Schwarzian action of the dual theory \cite{Grumiller:2017qao}. For this reason, we include its contribution in our analysis. 

Finally, one can construct models such that $(\partial_{u}X)/w \to 0$ but $(\partial_{u}X)^2/w$ diverges as $r \to \infty$. These tend to involve unusual potentials (for example, $U=0$ and $V \sim X^{\alpha}$ with $0 < \alpha < 1$) so we will not consider them further. Such models require additional terms in the derivative expansion of $L_\text{CT}$, which can be determined using the prescription above. In models where $(\partial_{u}X)/w$ remains finite as $r \to \infty$, all terms in the derivative expansion \eqref{eq:GeneralLct} have similar leading behavior and must be retained. And if $(\partial_{u}X)/w$ diverges in this limit then subsequent terms in the expansion are more divergent than previous terms. In that case the derivative expansion \eqref{eq:LctAnsatz} is not a useful organizing principle. Such models may be of interest, but they require a different approach to finding $L_\text{CT}$ and will not be considered here.

\section{Foliations of $\B$}
\label{app:geometry}

Let $\B$ be a two-dimensional surface with coordinates $x^{a}$ and Lorentzian metric $\gamma_{ab}$. 
\begin{gather}\label{eq:2dMetricAppendix}
    \gamma_{ab} dx^{a} dx^{b} = \V\,e^{2\beta}\,du^{2} + \W^{2}\,\big(d\phi - \U\,du\big)^{2} ~.
\end{gather}
The two-dimensional covariant derivative compatible with the metric is $D_{a}$, the volume element is $\sqrt{-\gamma} = e^{\beta}\,\W\,\sqrt{-\V}$, and the Ricci scalar is ${}^{2}\!R(\gamma)$.

Foliating $\B$ with curves of constant $u$, the future-directed timelike vector $\rho^{a}$ normal to these curves is 
\begin{gather}
    \rho^{a} = \frac{1}{e^{\beta}\sqrt{-\V}}\,\big( \delta^{a}{}_{u} + \U\,\delta^{a}{}_{\phi}\big) ~,
\end{gather}
with acceleration $\A^{a} = \rho^{b} D_{b} \rho^{a}$ given by
\begin{gather}
    \A^{a} = \frac{1}{\W^{2}}\,\partial_{\phi}\log\big(e^{\beta}\sqrt{-\V}\big) \, \delta^{a}{}_{\phi} ~.
\end{gather}
The extrinsic curvatures of these curves embedded in $\B$ is
\begin{gather}
    (K_c)_{ab} = \frac{1}{2}\,\big(D_{a} \rho_{b} + D_{b}\rho_{a} + \rho_a\,\A_b + \rho_b\,\A_a \big) ~.
\end{gather}
The trace $K_c = D_{a} \rho^{a}$ takes the simple form
\begin{gather}
    K_c = \frac{1}{\sqrt{-\gamma}}\,\Big(\partial_{u}\W + \partial_{\phi}(\W\,\U) \Big) ~.
\end{gather}
The constant-$u$ curves are one dimensional, so they have no intrinsic curvature and their the extrinsic curvature satisfies $K_c^{2} = (K_c)^{ab}\,(K_c)_{ab}$. As a result, the only Gauss-Codazzi identity for the two-dimensional curvature is
\begin{gather}
    {}^{2}\!R(\gamma) = D_{a}\big(2\,\rho^{a}\,K_c\big) - 2\,D_{a}\A^{a} ~.
\end{gather}
In the coordinates \eqref{eq:2dMetricAppendix}, and including a factor of the volume element, this can be written as
\begin{gather}
    \sqrt{-\gamma}\,{}^{2}\!R(\gamma) = \sqrt{-\gamma} \,D_{a}\big(2\,\rho^{a}\,K_c\big) - \partial_{\phi}\Big(\frac{2}{\W}\,\partial_{\phi}\big(e^{\beta}\,\sqrt{-\V}\big)\Big) ~.
\end{gather}
Using this to rewrite the $K_c$ corner term in the action \eqref{eq:3dActionWithCornersNewTerms}, the proposed action is
\begin{align}
    \Gamma = & \,\,  \frac1{2\kappa^{2}} \int_{M} \nts d^{3}x \sqrt{-g}\,\left(R+ \frac{2}{\ell^2} \right)  + \frac{1}{\kappa^2}\,\int_{\B}d^{2}x \sqrt{-\gamma} \left( K-\frac1{\ell} \right) - \frac{1}{\kappa^{2}} \,\int_{\partial \B}\!\! d^{1}x  \, \sqrt{\sigma} \\
    & + \frac{1}{\kappa^2}\,\int_{\B}d^{2}x \sqrt{-\gamma} \left(\f{\ell}2\,\,K_{c}^2 - \frac{\ell}{2}\,{}^{2}\!R(\gamma) \right)   ~.
\end{align} 
A total $\phi$-derivative in the integral over $\B$ has been discarded. The terms on the first line establish a well-defined variational principle with Dirichlet boundary conditions on the fields at $r \to \infty$, while the terms on $\B$ in the second line are an additional functional of the Dirichlet boundary data which must be included for the action to have a finite flat-space limit $\ell \to \infty$.

\section{Renormalization of the phase space in Fefferman--Graham gauge }
\label{app:3dFG}

In section \ref{sec1:covphasespace} we showed that the variation of the bulk action alone is sufficient to define a finite codimension-2 form at the asymptotic boundary that leads to integrable charges. This was applied to three dimensional gravity with a negative cosmological constant in section \ref{sec:3d}, working in Bondi-Weyl gauge. In this appendix we obtain similar results in Fefferman--Graham (FG) gauge \cite{fefferman1985conformal}. First, we show that components of the presymplectic potential parallel to the conformal boundary can be written as total (normal) derivatives. These shift the component normal to the conformal boundary by corner terms. The shifted presymplectic potential contains terms which diverge at the conformal boundary, but they are $\delta$-exact and hence do not contribute to the presymplectic current or codimension-2 form. A quick calculation shows that they are canceled in $\delta \Gamma$ by the usual boundary terms, including a non-trivial cancellation in the presence of corners.

\subsubsection*{Solution Space}

The line element in Fefferman-Graham gauge is 
\begin{equation}
\de s^2= \frac{\ell^2}{\rho^2} \de\rho^2 +\frac1{\rho^2} \gamma_{ab}(\rho,x^a)\de x^a \de x^b
\end{equation}
with $x^a=(t,\phi)$. The metric on a two-dimensional surface of constant $\rho$ is $\gamma_{ab}$, and the covariant derivative compatible with this metric is $D_{a}$. The conformal boundary is at $\rho \to 0$, and the outward-pointing, spacelike unit vector normal to a constant-$\rho$ surface is
\begin{gather}
    n_{\mu} = - \frac{\ell}{\rho}\,\delta_{\mu}{}^{\rho} \qquad n^{\mu} = - \frac{\rho}{\ell}\,\delta^{\mu}{}_{\rho} ~.
\end{gather}
Many of the calculations in this appendix involve quantities that diverge at $\rho \to 0$. We regulate these calculations by introducing a cut-off  $\rho = \epsilon$ with $0 < \epsilon \ll 1$, addressing any divergent terms, and then taking the limit $\epsilon \to 0$.

Partially on-shell fields satisfy $E^{\mu\nu}\,\delta g_{\mu\nu} = 0$ by fixing $\delta g_{\rho\rho} = \delta g_{\rho a} = 0$ and imposing the components $E^{ab}=0$ of the equations of motion. This fixes the $\rho$-dependence of $\gamma_{ab}$\,\footnote{In 3 dimensions, the expansion is finite and there are no subleading log terms consistent with the equations of motion $E^{ab}=0$. }  
\be
\gamma_{ab}= \gamma^\ms{(0)}_{ab}+ \gamma^\ms{(2)}_{ab} \rho^2 + \gamma^\ms{(4)}_{ab}\rho^4
\ee
with 
\begin{equation}
\gamma^\ms{(4)}_{ab}=\f14 \gamma^\ms{(2)}_{ac} \gamma_\ms{(0)}^{cd} \gamma^\ms{(2)}_{db} ~.
\end{equation}
The remaining components of the equations of motion, which we do not enforce, place constraints on the trace and covariant divergence of $\gamma^{\text{\tiny$(2)$}}_{ab}$. It is convenient to define 
\begin{equation}
    T_{ab}=\f1\ell \gamma^\ms{(2)}_{ab}+\frac{\ell}{2}\, \gamma^\ms{(0)}_{ab} \,\R \,, 
\end{equation}
where $\R$ is the Ricci scalar of the metric $\gamma^\ms{(0)}_{ab}$. Then the equations $E^{\rho\rho}=0 $ and $E^{\rho a}=0$ enforce respectively the conditions 
\begin{equation}
  T^{a}{}_{a}=\f{\ell}2 \,\R\,\,,\qquad  \D_{b}T^{ab}=0 ~.
\end{equation}
Indices are raised and lowered in these expressions with the metric $\gamma_{ab}^\ms{(0)}$, and $\D$ is the covariant derivative compatible with $\gamma_{ab}^\ms{(0)}$.

\subsubsection*{Renormalized symplectic potential}

Field variations that preserve the FG gauge satisfy
\begin{gather}
    \delta g_{\rho\rho} = \delta g_{\rho a} = 0 \qquad \delta g_{ab} = \frac{1}{\rho^{2}}\,\delta \gamma_{ab} ~.
\end{gather}
So the components of the presymplectic potential are
\begin{align}
    \Theta^{\rho} = &\,\, \frac{1}{2\kappa^{2}}\,\sqrt{-g}\,\Big(\nabla_{\nu}(\delta g)^{\rho \nu} - \nabla^{\rho}(\delta g)^{\nu}{}_{\nu}\Big) \\
    \Theta^{a} 
    = & \,\, \frac{1}{2\kappa^{2}}\frac{\ell}{\rho}\,\sqrt{-{\gamma}}\,\Big({D}_{b}(\delta {\gamma})^{ab} - 
{D}^{a}(\delta {\gamma})^{b}{}_{b}\Big) ~.
\end{align}
Only the leading part of $\Theta^{a}$ as $\rho \to 0$ will be relevant when we consider terms in the first variation of the action at $\rho = \epsilon$. This is
\begin{align}
    \Theta^{a} = \frac{1}{2\kappa^{2}}\,\frac{\ell}{\rho}\,\sqrt{-\gamma^\ms{(0)}}\,\Big(\D_{b}(\delta \gamma^\ms{(0)})^{ab} - \D^{a}(\delta \gamma^\ms{(0)})^{b}{}_{b} \Big) + \ldots ~.
\end{align}
The trailing $\ldots$ indicate terms that vanish as $\rho \to 0$. This can be written as a total $\rho$-derivative $\Theta^{a} = \partial_{\rho} Y^{a\rho}$ with
\begin{gather}
    Y^{a\rho} = \frac{1}{2\kappa^{2}}\,{\ell}\,\log\left(\frac{\rho}{\ell}\right)\,\sqrt{-\gamma^\ms{(0)}}\,\Big(\D_{b}(\delta \gamma^\ms{(0)})^{ab} - \D^{a}(\delta \gamma^\ms{(0)})^{b}{}_{b} \Big) +\mathcal O(\rho^2)
\end{gather}
Thus, the total derivative term in the variation of the bulk Lagrangian, for partially on-shell fields, is given by
\begin{gather}
    \delta L_{M} = \partial_{\mu}\Theta^{\mu} = \partial_{\rho}\left(\Theta^{\rho} + \partial_{a} Y^{a\rho} \right)=: \partial_\rho \Theta^\rho_{\text{ren}}
\end{gather}
with
\begin{align}
    \partial_{a} Y^{a\rho} = &\,\, \frac{1}{2\kappa^{2}}\,\ell\,\log\left(\frac{\rho}{\ell}\right)\,\sqrt{-\gamma^\ms{(0)}} \Big(\D_{a}\D_{b}(\delta \gamma^\ms{(0)})^{ab} - \D^{2}(\delta \gamma^\ms{(0)})^{b}{}_{b} \Big) +\mathcal O(\rho^2) \\ \label{eq:FGdY}
   = &\,\,\delta\left[  \frac{1}{2\kappa^{2}}\, \ell\log\left(\frac{\rho}{\ell}\right)\,\sqrt{-\gamma^\ms{(0)}} 
 \,\R\right] +\mathcal O(\rho^2) ~.
\end{align}
Combining this with $\Theta^{\rho}$, the shifted form of the presymplectic potential is
\begin{align}\label{FGsymplren}
    \Theta^\rho_{\text{ren}}&= -\frac{1}{2\kappa^{2}\,\ell}
\sqrt{-\gamma^\ms{(0)}}\,\Big(\gamma_\ms{(2)}^{ab} - \gamma_\ms{(0)}^{ab}\,\gamma^\ms{(2)}\Big) \delta \gamma^\ms{(0)}_{ab}\\ \label{FGsymplren2}
&+\frac{1}{\kappa^{2} }\,\delta\left[  \frac{\ell}{2}\log\left(\frac{\rho}{\ell}\right) \, \sqrt{-\gamma^\ms{(0)}} \,\R  +  \frac{1}{\ell}\frac1{\rho^2}\sqrt{-\gamma^\ms{(0)}} \left(1-\frac12  \gamma^\ms{(2)}\rho^2\right)\right] +\mathcal O(\rho^2)
\end{align}
Taking $\Theta^\rho_{\text{ren}}$ as the starting point leads to a finite (as $\rho \to 0$) codimension-2 form that yields charges. The $\delta$-exact terms in the second line identify boundary terms, including the holographic renormalization counterterms, needed for a well-defined variational principle \cite{deHaro:2000xn , Bianchi:2001kw}.

\subsubsection*{Renormalized action}

The full action for Einstein gravity with AdS$_3$ boundary conditions requires various boundary terms on the regulating surface $\rho=\epsilon$.
\begin{equation}\label{eq:FGBndyTerms}
    L_b= L_{GHY}+L_{CT} \vphantom{\Big|} ~.
\end{equation}
The first of these is the Gibbons-Hawking-York term
\begin{gather}
    L_{GHY} = \frac{1}{\kappa^2}\frac{\sqrt{-\gamma}}{\rho^2} K ~,   
\end{gather}
with $K = \nabla_{\mu} n^{\mu}$ the trace of the extrinsic curvature. The second is a boundary counterterm, which includes a log term proportional to the Ricci scalar for $\gamma^\ms{(0)}_{ab}$
\begin{gather}\label{eq:FGCT}
    L_{CT} = \frac{1}{\kappa^{2}}\,\frac{\sqrt{-\gamma}}{\rho^2}\,\Big(- \frac{1}{\ell} + \frac{\ell}{2}\,\R\,{\rho^{2}}\log\Big(\frac{\rho}{\ell}\Big) \Big) ~.
\end{gather}
One can easily check that the variation of these boundary terms cancels the $\rho = \epsilon$ contribution to $\delta\Gamma$ coming from the $\delta$-exact term in \eqref{FGsymplren2}. Accounting for both the bulk and boundary terms, the terms on $\rho = \epsilon$ in the first variation of the action are finite as $\epsilon \to 0$  
\begin{align}
    \delta \Gamma 
    = &\,\, \frac{1}{2\kappa^{2}}\,\int_{\B} d^{2}x\,\sqrt{-\gamma^\ms{(0)}}\,\frac{1}{\ell}\,\Big(\gamma_\ms{(2)}^{ab} - \gamma_\ms{(0)}^{ab}\,\gamma^\ms{(2)}\Big) \delta \gamma^\ms{(0)}_{ab}  \\
    = &\,\, \frac{1}{2\kappa^{2}}\,\int_{\B} d^{2}x\,\sqrt{-\gamma^\ms{(0)}}\,\Big( T^{ab} - \gamma_\ms{(0)}^{ab}\,T^{c}{}_{c} + \frac{\ell^{2}}{2}\,\gamma_\ms{(0)}^{ab}\,\R \Big) \delta \gamma^\ms{(0)}_{ab}  ~.    
\end{align}
When the $E^{\rho\rho} = 0$ constraint is enforced, the last two terms cancel and the integrand reduces to the boundary stress tensor times the metric variation -- precisely the expected result. The main observation here is that the corner term \eqref{eq:FGdY}, which comes from expressing $\Theta^{a}$ as a total $\rho$-derivative, precisely cancels the variation of the log-divergent part of the counterterm \eqref{eq:FGCT}. This detail is not important when $\partial\B = 0$, since the log divergence in $\delta L_{CT}$ is a total derivative. But when $\B$ has corners this cancellation is necessary.

\providecommand{\href}[2]{#2}\begingroup\raggedright\endgroup

\end{document}